\documentclass[12pt]{article}
\usepackage{amsfonts,graphicx,amsmath,amsthm,amssymb,epsfig}
\usepackage{float}
\allowdisplaybreaks
\begin{document}

\title{\bf Effects of Non-minimal Matter-geometry Coupling on Embedding Class-one Anisotropic Solutions}
\author{M. Sharif \thanks{msharif.math@pu.edu.pk} and T. Naseer \thanks{tayyabnaseer48@yahoo.com}\\
Department of Mathematics, University of the Punjab,\\
Quaid-i-Azam Campus, Lahore-54590, Pakistan.}

\date{}
\maketitle

\begin{abstract}
This paper investigates some particular anisotropic star models in
$f(\mathcal{R},\mathcal{T},\mathcal{Q})$ gravity, where
$\mathcal{Q}=\mathcal{R}_{\omega\alpha}\mathcal{T}^{\omega\alpha}$.
We adopt a standard model
$f(\mathcal{R},\mathcal{T},\mathcal{Q})=\mathcal{R}+\varpi\mathcal{Q}$,
where $\varpi$ indicates a coupling constant. We take spherically
symmetric spacetime and develop solutions to the modified field
equations corresponding to different choices of the matter
Lagrangian by applying `embedding class-one' scheme. For this
purpose, we utilize $\mathbb{MIT}$ bag model equation of state and
investigate some physical aspects of compact models such as RXJ
1856-37,~4U 1820-30,~Cen X-3,~SAX J 1808.4-3658 and Her X-I. We use
masses and radii of these stars and employ the vanishing radial
pressure condition at the boundary to calculate the value of their
respective bag constant $\mathfrak{B_c}$. Further, we fix
$\varpi=\pm4$ to analyze the behavior of resulting state variables,
anisotropy, mass, compactness, surface redshift as well as energy
bounds through graphical interpretation for each star model. Two
different physical tests are performed to check the stability of the
developed solutions. We conclude that $\varpi=-4$ is more suitable
choice for the considered modified model to obtain stable structures
of the compact bodies.
\end{abstract}

\section{Introduction}

Although General Relativity ($\mathbb{GR}$) has made tremendous
achievements in elucidating many unknown components of our universe,
it is insufficient to investigate cosmic structure at vast scale. In
the recent era, several modifications to $\mathbb{GR}$ have been
postulated to disclose the perplexing issues which are associated
with the cosmic evolution such as its rapid expansion and dark
matter etc. This expansion results in the existence of an obscure
form of force, named as dark energy having immensely large negative
pressure. Therefore, the modified theories are viewed as remarkably
significant in exposing the cosmic mysterious features. Firstly, the
geometric part of Einstein-Hilbert action was modified to obtain
$f(\mathcal{R})$ theory which is straightforward generalization to
$\mathbb{GR}$ due to the insertion of generic function of the Ricci
scalar in place of $\mathcal{R}$. Various authors \cite{2}-\cite{2d}
have studied the celestial structures in this theory and analyzed
their feasibility through different schemes. Multiple forms of
$f(\mathcal{R})$ gravity have been considered to study several
cosmological problems, i.e., late-time evolution of the universe
\cite{3,3b}, the inflationary era \cite{4} and background of cosmic
expansion \cite{5,5a}.

Bertolami \emph{et al.} \cite{10} analyzed the influence of
matter-geometry coupling on stellar systems in $f(\mathcal{R})$
scenario for the very first time by adopting the matter Lagrangian
in form of $\mathcal{R}$ and $\mathbb{L}_{m}$. Many researchers have
been prompted by such interaction, as a result of which they payed
their concentration in investigating cosmic rapid expansion. A
couple of years ago, several modified gravitational theories were
proposed comprising an arbitrary coupling in some general manner at
the action level which ultimately turns out to be a subject of great
importance for astrophysicists. The $f(\mathcal{R},\mathcal{T})$
gravity which encompasses such interaction is contemplated by Harko
\emph{et al.} \cite{20}, in which $\mathcal{T}$ demonstrates trace
of the energy-momentum tensor $(\mathbb{EMT})$. The modified
functional forms involving $\mathcal{T}$ give rise to the
non-conserved $\mathbb{EMT}$ unlike $\mathbb{GR}$ and
$f(\mathcal{R})$ theories. Numerous scientists \cite{21}-\cite{21f}
investigated different massive structures in
$f(\mathcal{R},\mathcal{T})$ framework and found that this gravity
yields several fascinating astrophysical outcomes. To get better
understanding of cosmic inflationary era, an even more complex
functional $f(\mathcal{R},\mathcal{T},\mathcal{Q})$ is suggested by
Haghani \emph{et al.} \cite{22}, where $\mathcal{Q}$ indicates the
contraction of the Ricci tensor and $\mathbb{EMT}$ (i.e.,
$\mathcal{Q}\equiv
\mathcal{R}_{\omega\alpha}\mathcal{T}^{\omega\alpha}$). They deemed
certain mathematical models to study their physical feasibility and
employed Lagrange multiplier method to obtain conservation of the
$\mathbb{EMT}$.

The development of this modified gravity was premised on the
insertion of the factor $\mathcal{Q}$ which ensures the presence of
strong non-minimal matter-geometry coupling in self-gravitating
systems. The modification in the Einstein-Hilbert action may help in
explaining the role of dark energy and dark matter, without
resorting to exotic fluid distribution. Some other extensions to
$\mathbb{GR}$ like $f(\mathcal{R},\mathbb{L}_m)$ and
$f(\mathcal{R},\mathcal{T})$ gravitational theories also comprise
the matter Lagrangian involving such arbitrary interaction but we
cannot consider their functionals as the most generalized form that
provide proper understanding to the influence of coupling on
self-gravitating objects in some scenarios. It must be noted here
that the factor
$\mathcal{R}_{\omega\alpha}\mathcal{T}^{\omega\alpha}$ could explain
the impact of non-minimal interaction in the situation where
$f(\mathcal{R},\mathcal{T})$ theory breaks down to achieve such
results. In particular, $f(\mathcal{R},\mathcal{T})$ fails to uphold
the non-minimal coupling for the case when $\mathcal{T}=0$ inside
the configuration, however, this phenomenon can be explained by
$f(\mathcal{R},\mathcal{T},\mathcal{Q})$ gravity. Due to the
non-conservation of energy-momentum tensor in this theory, an
additional force is present due to which the motion of test
particles in geodesic path comes to an end. This force also helps to
elucidate the galactic rotation curves. Sharif and Zubair \cite{22a}
assumed two models such as $\mathcal{R}+\varpi\mathcal{Q}$ and
$\mathcal{R}(1+\varpi\mathcal{Q})$ to study thermodynamical laws for
black holes with different choices of matter Lagrangian like
$\mathbb{L}_m=\rho$ as well as $-p$ and obtained their viability
constraints. They also explored energy conditions corresponding to
the above models and concluded that weak energy conditions are
satisfied only when $\varpi>0$ \cite{22b}.

Odintsov and S\'{a}ez-G\'{o}mez \cite{23} studied various
cosmological solutions and the occurrence of $\Lambda$CDM model in
$f(\mathcal{R},\mathcal{T},\mathcal{Q})$ gravity. They also
discussed the issue related to matter instability and figured out
that this theory may permit to generate pure de Sitter universe
subject to the usability of non-constant fluids. An important
requirement usually demanded by extended theories consists of the
avoidance of the Ostrogradski as well as Dolgov-Kawasaki
instability. The later issue has recently been addressed for this
theory \cite{22,23}. Ayuso \emph{et al.} \cite{24} showed that
conformal as well as strong non-minimal matter-geometry couplings
oftenly produce higher-order equations of motion and illustrated
this phenomenon by adopting certain appropriate scalar as well as
massive vector fields. It is found that ghost modes are generally
present in these theories due to the coupling
$\mathcal{R}_{\omega\alpha}\mathcal{T}^{\omega\alpha}$ and that its
avoidance considerably restricts the allowed form for the functional
$f$. Different models in this theory have been studied, one of them
is the $f(\mathcal{R},\mathcal{T},\mathcal{Q})=\beta(\mathcal{R})^n
+\varpi(\mathcal{Q})^m$, where $\beta$ and $\varpi$ are real-valued
coupling constants. It is shown that this theory will be free of
Ostrogradski instabilities for $n=1=m$.

Baffou \emph{et al.} \cite{25} obtained Friedmann equations and
developed the stability of this theory for two different models by
calculating numerical solution of the perturbation functions as well
as power-law and de Sitter solutions. Sharif and Waseem \cite{25a}
calculated the solutions corresponding to the isotropic/anisotropic
configurations by choosing different matter Lagrangian and discussed
their stability through various approaches. Yousaf \emph{et al.}
\cite{26}-\cite{26e} studied the structural evolution of spherical
and cylindrical celestial bodies with the help of modified structure
scalars which came from orthogonal decomposition of the Riemann
tensor. We have utilized the decoupling technique to get physically
acceptable charged/uncharged solutions to the
$f(\mathcal{R},\mathcal{T},\mathcal{Q})$ field equations
\cite{27,27a}.

Stars are acknowledged as indispensable components of our galaxy
among the plenty of unfathomable constituents of the universe. The
structural formation of such objects encouraged many astrophysicists
that they would pay attention on the study of their developmental
phases. Neutron stars gained much attention amongst all compact
bodies in virtue of their intriguing properties. The mass of a
neutron star is approximately 1 to 3 times solar masses
($M_{\bigodot}$) and its core contains newly formed neutrons which
help to produce degeneracy pressure to counterbalance the
gravitational force and resist that neutron to further collapse. The
first ever neutron star was predicted in 1934 \cite{28}, but
observationally, it was confirmed later. This is due to the fact
that neutron stars do not release sufficient radiations and are
mostly indiscernible. Another highly dense object between neutron
star and black hole is the quark star whose interior is filled with
up, down and strange quark matter. Numerous research has been done
on the study of formation of such hypothetical structures
\cite{29a}-\cite{29h}.

At present time, the study of compact bodies whose interiors contain
anisotropic matter configuration have become a persuasive subject of
research for numerous astronomers. According to Herrera' observation
\cite{29i}, a celestial object comprising nuclear density in its
core much lower than the mass density should be owned by anisotropic
fluid. Herrera and Santos \cite{30} discussed the self-gravitating
structures and analyzed impressive effects of anisotropy on those
bodies. Harko and Mak \cite{31} inspected the feasibility of
anisotropic solutions corresponding to static relativistic objects.
Hossein \emph{et al.} \cite{32} considered Krori-Barua solution to
analyze stability of the anisotropic massive systems by including
the impact of cosmological constant $\Lambda$. Kalam \emph{et al.}
\cite{32a} constructed solutions of gravitational equations of
motion corresponding to various neutron stars and found them viable
as well as stable. Paul and Deb \cite{32b} developed certain
feasible solutions for compact stars which were considered in
hydrostatic equilibrium.

The matter distribution inside quark bodies contains physical
variables such as energy density and pressure, thus in this regard,
the $\mathbb{MIT}$ bag model equation of state
($\mathbb{E}$o$\mathbb{S}$) is viewed as an effective tool which
interrelates these quantities \cite{33}. It has also been observed
that this model can efficiently describe the compactness of various
structures such as PSR 0943+10, 4U 1820-30, 4U 1728-34, RXJ
185635-3754, Her X-1 and SAX J 1808.4-3658, etc., while the neutron
star $\mathbb{E}$o$\mathbb{S}$ flunks in this context \cite{33a}.
Generally, a vacuum involves true and false states whose discrepency
can be determined by the bag constant ($\mathfrak{B_c}$) and its
increasing value results in lowering quark pressure. Several
researchers \cite{33b}-\cite{34aa} analyzed different quark stars
and their internal formation with the help of $\mathbb{MIT}$ bag
model. Demorest \emph{et al.} \cite{34b} quantified the mass of a
quark star (namely, PSR J1614-2230) and concluded that only this
model can support such kind of heavily objects. The mass of various
compact structures has been measured by Rahaman \emph{et al.}
\cite{35} by employing an interpolating technique and they also
studied some physical aspects of a star candidate having radius as
9.9km.

Several techniques have been used in literature to formulate
solution to the field equations such as the solution of metric
potentials or by making use of a particular
$\mathbb{E}$o$\mathbb{S}$. An embedding class-one technique is one
of them which states that $(n-2)$-dimensional space can be embedded
into an $(n-1)$-dimensional space. Bhar \emph{et al.} \cite{36}
utilized this scheme along with a new metric coefficient to find
physically feasible solutions for anisotropic systems. Maurya
\emph{et al.} \cite{37,37a} calculated the embedding class-one
solution and analyzed its stability as well as effects of anisotropy
on relativistic stars. By suggesting a particular metric function,
Singh \emph{et al.} \cite{37b} formed singularity-free solution for
spherical geometry with the help of this technique. This work has
been extended by Sharif and his collaborators \cite{38}-\cite{38g}.
They found stable as well as viable solutions in different theories
like $f(\mathcal{R},\mathcal{T})$, Brans-Dicke and $f(\mathcal{G})$
gravity.

In this paper, we analyze physical feasibility of two solutions to
the modified field equations corresponding to different forms of
matter Lagrangian in
$f(\mathcal{R},\mathcal{T},\mathcal{R}_{\omega\alpha}\mathcal{T}^{\omega\alpha})$
theory. The following lines help to understand that how the paper is
structured. In the next section, we construct the field equations in
modified gravity for a particular model ($\mathcal{R}+\varpi
\mathcal{R}_{\omega\alpha}\mathcal{T}^{\omega\alpha}$), where we fix
$\varpi=\pm4$. Further, we assume $\mathbb{MIT}$ bag model
$\mathbb{E}$o$\mathbb{S}$ to solve the field equations, take one
metric potential and employ embedding class-one condition to find
the other. Section 3 calculates the four unknowns $(W,X,Y,Z)$ at the
hypersurface. Various physical characteristics of compact stars are
analyzed through graphical interpretation in section 4. Lastly, the
concluded remarks are presented in section 5.

\section{The $f(\mathcal{R},\mathcal{T},\mathcal{R}_{\omega\alpha}\mathcal{T}^{\omega\alpha})$ Gravity}

The modified form of Einstein-Hilbert action in terms of complex
analytical functional
$f(\mathcal{R},\mathcal{T},\mathcal{R}_{\omega\alpha}\mathcal{T}^{\omega\alpha})$
(with $\kappa=8\pi$) is given as \cite{23}
\begin{equation}\label{g1}
S_{f(\mathcal{R},\mathcal{T},\mathcal{R}_{\omega\alpha}\mathcal{T}^{\omega\alpha})}=\int\sqrt{-g}
\left[\frac{f(\mathcal{R},\mathcal{T},\mathcal{R}_{\omega\alpha}\mathcal{T}^{\omega\alpha})}{16\pi}
+\mathbb{L}_{m}\right]d^{4}x,
\end{equation}
where $\mathbb{L}_{m}$ serves as the matter Lagrangian. After
applying the variational principle on the action \eqref{g1}, we
obtain the field equations as
\begin{equation}\label{g2}
\mathcal{G}_{\omega\alpha}=\mathcal{T}_{\omega\alpha}^{(eff)}=8\pi\bigg(\frac{\mathcal{T}_{\omega\alpha}}{f_{\mathcal{R}}
-\mathbb{L}_{m}f_{\mathcal{Q}}}+\mathcal{T}_{\omega\alpha}^{(\mathcal{C})}\bigg),
\end{equation}
which describe matter in terms of spacetime. Here,
$\mathcal{G}_{\omega\alpha}$ represents the geometrical structure
and $\mathcal{T}_{\omega\alpha}^{(eff)}$ is the $\mathbb{EMT}$ in
modified gravity which involves state variables along with their
derivatives. Thus the sector
$\mathcal{T}_{\omega\alpha}^{(\mathcal{C})}$ appearing due to the
insertion of additional term
$\mathcal{R}_{\omega\alpha}\mathcal{T}^{\omega\alpha}$ in the action
\eqref{g1} is given as
\begin{eqnarray}\nonumber
\mathcal{T}_{\omega\alpha}^{(\mathcal{C})}&=&\frac{1}{8\pi\left(f_{\mathcal{R}}-\mathbb{L}_{m}f_{\mathcal{Q}}\right)}
\left[\bigg(f_{\mathcal{T}}+\frac{1}{2}\mathcal{R}f_{\mathcal{Q}}\bigg)\mathcal{T}_{\omega\alpha}
+\left\{\frac{\mathcal{R}}{2}(\frac{f}{\mathcal{R}}-f_{\mathcal{R}})-\mathbb{L}_{m}f_{\mathcal{T}}\right.\right.\\\nonumber
&-&\left.\frac{1}{2}\nabla_{\sigma}\nabla_{\xi}(f_{\mathcal{Q}}\mathcal{T}^{\sigma\xi})\right\}g_{\omega\alpha}
-\frac{1}{2}\Box(f_{\mathcal{Q}}\mathcal{T}_{\omega\alpha})-(g_{\omega\alpha}\Box-
\nabla_{\omega}\nabla_{\alpha})f_{\mathcal{R}}\\\label{g4}
&-&2f_{\mathcal{Q}}\mathcal{R}_{\sigma(\omega}\mathcal{T}_{\alpha)}^{\sigma}
+\nabla_{\sigma}\nabla_{(\omega}[\mathcal{T}_{\alpha)}^{\sigma}f_{\mathcal{Q}}]+2(f_{\mathcal{Q}}\mathcal{R}^{\sigma\xi}
+\left.f_{\mathcal{T}}g^{\sigma\xi})\frac{\partial^2\mathbb{L}_{m}}{\partial
g^{\omega\alpha}\partial g^{\sigma\xi}}\right],
\end{eqnarray}
where $f_{\mathcal{R}}=\frac{\partial
f(\mathcal{R},\mathcal{T},\mathcal{Q})}{\partial
\mathcal{R}},~f_{\mathcal{T}}=\frac{\partial
f(\mathcal{R},\mathcal{T},\mathcal{Q})}{\partial \mathcal{T}}$ and
$f_{\mathcal{Q}}=\frac{\partial
f(\mathcal{R},\mathcal{T},\mathcal{Q})}{\partial \mathcal{Q}}$.
Also, $\nabla_\omega$ and $\Box\equiv
\frac{1}{\sqrt{-g}}\partial_\omega\big(\sqrt{-g}g^{\omega\alpha}\partial_{\alpha}\big)$
represent covariant derivative and the D'Alambert operator,
respectively. The attractive nature of this modified gravity needs
to satisfy the following constraint
\begin{equation}\nonumber
\frac{8\pi+f_{\mathcal{T}}+\frac{1}{2}\mathcal{R}f_{\mathcal{Q}}}{f_{\mathcal{R}}-\mathbb{L}_{m}f_{\mathcal{Q}}}>0.
\end{equation}

The modified field equations involve explicit form of the matter
Lagrangian, thus the corresponding dynamics can be studied by taking
some particular form of this Lagrangian. Different choices of the
matter Lagrangian for a perfect fluid have been analyzed in
literature. These include $\mathbb{L}_m=\pm{P}$ \cite{39},
$\mathbb{L}_m=\pm{\mu}$ \cite{39a} and $\mathbb{L}_m=\mathcal{T}$
\cite{40}, where $P$ and $\mu$ are isotropic pressure and energy
density, respectively. The sign depends on the signature of the
chosen metric. As we have considered anisotropic matter distribution
in the interior of quark stars, thus the Lagrangian $\mathbb{L}_m$
in terms of pressure $P$ can now be taken as $P_{r}$ (radial
pressure) and $P_{\bot}$ (tangential pressure). Finally, we have
three choices of the matter Lagrangian in this case as $\mu,~P_{r}$
and $P_{\bot}$. According to the signatures ($-,+,+,+$), one can
take $\mathbb{L}_m=-\mu,~P_{r}$ and $P_{\bot}$. Here, we adopt first
two forms (which has been extensively employed) to obtain the
solutions of modified field equations and to analyze the effects of
strong matter-geometry coupling on them. These choices lead to
$\frac{\partial^2\mathbb{L}_{m}} {\partial g^{\omega\alpha}\partial
g^{\sigma\xi}}=0$ \cite{22}. The choice $\mathbb{L}_m=P_{\bot}$ can
also be taken which may produce acceptable results in this theory.

The covariant divergence of the effective $\mathbb{EMT}$ has the
form
\begin{align}\nonumber
\nabla^\omega \mathcal{T}_{\omega\alpha}^{(eff)}=0,
\end{align}
which consequently leads to
\begin{align}\nonumber
\nabla^\omega
\mathcal{T}_{\omega\alpha}&=\frac{2}{2f_\mathcal{T}+\mathcal{R}f_\mathcal{Q}+16\pi}\bigg[\nabla_\omega\big(f_\mathcal{Q}\mathcal{R}^{\sigma\omega}
\mathcal{T}_{\sigma\alpha}\big)-\mathcal{G}_{\omega\alpha}\nabla^\omega\big(f_\mathcal{Q}\mathbb{L}_m\big)
-\frac{1}{2}\nabla_\alpha\mathcal{T}^{\sigma\xi}\\\label{g4a}
&\times\big(f_\mathcal{T}g_{\sigma\xi}+f_\mathcal{Q}\mathcal{R}_{\sigma\xi}\big)+\nabla_\alpha\big(\mathbb{L}_mf_\mathcal{T}\big)
-\frac{1}{2}\big\{\nabla^{\omega}(\mathcal{R}f_{\mathcal{Q}})+2\nabla^{\omega}f_{\mathcal{T}}\big\}\mathcal{T}_{\omega\alpha}\bigg].
\end{align}

The $\mathbb{EMT}$ is mainly used to describe the distribution of
matter in the interior of astronomical systems and its each non-null
component is related to a particular physical property of the
structure. The anisotropy (occurs when the pressure in radial and
tangential directions are different) in celestial objects is
considered as a valuable factor in examining the evolution of their
structures. Our cosmos contains a number of massive objects and most
of them are considered to be interlinked with anisotropic
distribution, therefore this factor has important implications for
stellar models in their evolutionary phases. Hence, we assume the
following $\mathbb{EMT}$ representing anisotropic configuration as
\begin{equation}\label{g5}
\mathcal{T}_{\omega\alpha}=(\mu+P_\bot) \mathcal{K}_{\omega}
\mathcal{K}_{\alpha}+P_\bot
g_{\omega\alpha}+\left(P_r-P_\bot\right)\mathcal{W}_\omega\mathcal{W}_\alpha,
\end{equation}
where the quantities $\mathcal{W}_{\omega}$ and $\mathcal{K}_\omega$
indicate the four-vector and the four-velocity, respectively. The
trace of $f(\mathcal{R},\mathcal{T},\mathcal{Q})$ field equations
become as
\begin{align}\nonumber
&3\nabla^{\sigma}\nabla_{\sigma}
f_\mathcal{R}-\mathcal{R}\left(\frac{\mathcal{T}}{2}f_\mathcal{Q}-f_\mathcal{R}\right)-\mathcal{T}(8\pi+f_\mathcal{T})+\frac{1}{2}
\nabla^{\sigma}\nabla_{\sigma}(f_\mathcal{Q}\mathcal{T})\\\nonumber
&-2f+\nabla_\omega\nabla_\sigma(f_\mathcal{Q}\mathcal{T}^{\omega\sigma})+(\mathcal{R}f_\mathcal{Q}
+4f_\mathcal{T})\mathbb{L}_m+2\mathcal{R}_{\omega\sigma}\mathcal{T}^{\omega\sigma}f_\mathcal{Q}\\\nonumber
&-2g^{\alpha\xi} \frac{\partial^2\mathbb{L}_m}{\partial
g^{\alpha\xi}\partial
g^{\omega\sigma}}\left(f_\mathcal{T}g^{\omega\sigma}+f_\mathcal{Q}R^{\omega\sigma}\right)=0.
\end{align}
The insertion of $\mathcal{Q}=0$ in the overhead equation vanishes
the effects of strong non-minimal coupling in the interior of
stellar object and produces $f(\mathcal{R},\mathcal{T})$ theory,
which further reduces to the $f(\mathcal{R})$ gravity after
implementing the vacuum scenario.

We assume spherical spacetime to discuss the internal matter
distribution of compact stars as
\begin{equation}\label{g6}
ds^2=-e^{\rho} dt^2+e^{\zeta} dr^2+r^2d\theta^2+r^2\sin^2\theta
d\psi^2,
\end{equation}
where $\rho=\rho(r)$ and $\zeta=\zeta(r)$. This produces four-vector
and four-velocity in comoving coordinates as
\begin{equation}\label{g7}
\mathcal{W}^\omega=\delta^\omega_1 e^{\frac{-\zeta}{2}}, \quad
\mathcal{K}^\omega=\delta^\omega_0 e^{\frac{-\rho}{2}},
\end{equation}
which must satisfy $\mathcal{K}^\omega \mathcal{K}_{\omega}=-1$ and
$\mathcal{W}^\omega \mathcal{K}_{\omega}=0$, as we consider the
signatures of geometry \eqref{g6} $(-,+,+,+)$. Our universe is
currently passing through accelerating expansion phase and consists
of numerous stars existing in non-linear regime, but the study of
their linear behavior may provide better understanding about the
structural formation of these massive bodies. In contrast to
$f(\mathcal{R},\mathbb{L}_m)$ and $f(\mathcal{R},\mathcal{T})$
theories, the factor
$\mathcal{R}_{\omega\alpha}\mathcal{T}^{\omega\alpha}$ is
responsible to make $f(\mathcal{R},\mathcal{T},\mathcal{Q})$ gravity
much complicated. Haghani \emph{et al.} \cite{22} discussed
cosmological applications of three different models in this
framework, i.e.,
$\mathcal{R}+\varpi\mathcal{Q},~\mathcal{R}(1+\varpi\mathcal{Q})$
and $\mathcal{R}+\beta\sqrt{\mid\mathcal{T}\mid}+\varpi\mathcal{Q}$,
where $\varpi$ and $\beta$ are arbitrary coupling constants. They
analyzed the evolution and dynamics of the universe for the above
models with and without energy conservation. We thus consider the
following model to analyze five different star candidates as
\begin{equation}\label{g61}
f(\mathcal{R},\mathcal{T},\mathcal{R}_{\omega\alpha}\mathcal{T}^{\omega\alpha})=f_1(\mathcal{R})+
f_2(\mathcal{R}_{\omega\alpha}\mathcal{T}^{\omega\alpha})=\mathcal{R}+\varpi
\mathcal{R}_{\omega\alpha}\mathcal{T}^{\omega\alpha}.
\end{equation}
In this case, if $\varpi>0$, the solution has an oscillatory
behavior with alternating expanding and collapsing phases. For
$\varpi<0$, the scale factor of the universe has a hyperbolic
cosine-type dependence. Since $\varpi$ is any real-valued coupling
constant, thus one can take any value (positive or negative) to
check whether the corresponding solution is physically acceptable or
not. In this regard, we choose it as $\varpi=\pm4$.

It is interesting to stress that physical feasibility of different
gravity models can be achieved by taking the value of coupling
parameter within its observed range. The model \eqref{g61} has been
utilized in several investigations based on the stability and
viability of various isotropic and anisotropic configured stars
\cite{22a,22b,25a}. In this case, $\mathcal{Q}$ becomes
\begin{eqnarray}\nonumber
\mathcal{Q}&=&e^{-\zeta}\bigg[\frac{\mu}{4}\left(\rho'^2-\rho'\zeta'+2\rho''+\frac{4\rho'}{r}\right)-\frac{P_r}{4}\left(\rho'^2-\rho'\zeta'
+2\rho''+\frac{4\zeta'}{r}\right)\\\nonumber &+&P_\bot
\left(\frac{\zeta'}{r}-\frac{\rho'}{r}+\frac{2e^\zeta}{r^2}-\frac{2}{r^2}\right)\bigg].
\end{eqnarray}
Here, $'=\frac{\partial}{\partial r}$. We use Eqs.\eqref{g2} and
\eqref{g4} together along with the model \eqref{g61} to construct
modified field equations as
\begin{eqnarray}\nonumber
\mathcal{G}_{\omega\alpha}&=&\frac{\varpi}{1-\varpi \mathbb{L}_m}
\bigg[\left(\frac{8\pi}{\varpi}+\frac{1}{2}\mathcal{R}\right)\mathcal{T}_{\omega\alpha}+\frac{1}{2}\left\{\mathcal{Q}
-\nabla_{\sigma}\nabla_{\xi}\mathcal{T}^{\sigma\xi}\right\}g_{\omega\alpha}
-2\mathcal{R}_{\sigma(\omega}\mathcal{T}_{\alpha)}^{\sigma}\\\label{g7a}
&-&\frac{1}{2}\Box\mathcal{T}_{\omega\alpha}+\nabla_{\sigma}\nabla_{(\omega}\mathcal{T}_{\alpha)}^{\sigma}+2\mathcal{R}^{\sigma\xi}
\frac{\partial^2\mathbb{L}_{m}}{\partial g^{\omega\alpha}\partial
g^{\sigma\xi}}\bigg],
\end{eqnarray}
and the covariant divergence \eqref{g4a} becomes
\begin{eqnarray}\nonumber
\nabla^\omega
\mathcal{T}_{\omega\alpha}&=&\frac{2\varpi}{\varpi\mathcal{R}+16\pi}\bigg[\nabla_\omega(\mathcal{R}^{\sigma\omega}\mathcal{T}_{\sigma\alpha})
-\frac{1}{2}\mathcal{R}_{\sigma\xi}\nabla_\alpha\mathcal{T}^{\sigma\xi}-\frac{1}{2}
\mathcal{T}_{\omega\alpha}\nabla^\omega\mathcal{R}-\mathcal{G}_{\omega\alpha}\\\label{g7b}
&\times&\nabla^\omega\big(\mathbb{L}_m\big)\bigg].
\end{eqnarray}
The non-zero components of field equations \eqref{g7a} for the fluid
distribution \eqref{g5} and $\mathbb{L}_m=-\mu$ become
\begin{align}\nonumber
8\pi\mu&=e^{-\zeta}\bigg[\frac{\zeta'}{r}+\frac{e^\zeta}{r^2}-\frac{1}{r^2}+\varpi\bigg\{\mu\bigg(\frac{3\rho'\zeta'}{8}-\frac{\rho'^2}{8}
+\frac{\zeta'}{r}+\frac{e^\zeta}{r^2}-\frac{1}{r^2}-\frac{3\rho''}{4}\\\nonumber
&-\frac{3\rho'}{2r}\bigg)-\mu'\bigg(\frac{\zeta'}{4}-\frac{1}{r}-\rho'\bigg)+\frac{\mu''}{2}+P_r\bigg(\frac{\rho'\zeta'}{8}
-\frac{\rho'^2}{8}-\frac{\rho''}{4}+\frac{\zeta'}{2r}+\frac{\zeta''}{2}\\\label{g8}
&-\frac{3\zeta'^2}{4}\bigg)+\frac{5\zeta'P'_r}{4}-\frac{P''_r}{2}+P_\bot\bigg(\frac{\zeta'}{2r}-\frac{\rho'}{2r}
+\frac{3e^\zeta}{r^2}-\frac{1}{r^2}\bigg)-\frac{P'_\bot}{r}\bigg\}\bigg],\\\nonumber
8\pi
P_r&=e^{-\zeta}\bigg[\frac{\rho'}{r}-\frac{e^\zeta}{r^2}+\frac{1}{r^2}+\varpi\bigg\{\mu\bigg(\frac{\rho'\zeta'}{8}+\frac{\rho'^2}{8}
-\frac{\rho''}{4}-\frac{\rho'}{2r}\bigg)-\frac{\rho'\mu'}{4}-P_r\\\nonumber
&\times\bigg(\frac{5\rho'^2}{8}-\frac{7\rho'\zeta'}{8}+\frac{5\rho''}{4}-\frac{7\zeta'}{2r}+\frac{\rho'}{r}-\zeta'^2
-\frac{e^\zeta}{r^2}+\frac{1}{r^2}\bigg)+P'_r\bigg(\frac{\rho'}{4}+\frac{1}{r}\bigg)\\\label{g8a}
&-P_\bot\bigg(\frac{\zeta'}{2r}-\frac{\rho'}{2r}+\frac{3e^\zeta}{r^2}
-\frac{1}{r^2}\bigg)+\frac{P'_\bot}{r}\bigg\}\bigg],\\\nonumber 8\pi
P_\bot&=e^{-\zeta}\bigg[\frac{\rho'^2}{4}-\frac{\rho'\zeta'}{4}+\frac{\rho''}{2}-\frac{\zeta'}{2r}+\frac{\rho'}{2r}
+\varpi\bigg\{\mu\bigg(\frac{\rho'^2}{8}+\frac{\rho'\zeta'}{8}-\frac{\rho''}{4}-\frac{\rho'}{2r}\bigg)\\\nonumber
&-\frac{\rho'\mu'}{4}-P_r\bigg(\frac{\rho'\zeta'}{8}-\frac{\rho'^2}{8}-\frac{\rho''}{4}+\frac{\zeta'}{2r}+\frac{\zeta''}{2}
-\frac{3\zeta'^2}{4}\bigg)-\frac{5\zeta'P'_r}{4}+\frac{P''_r}{2}\\\label{g8b}
&-P_\bot\bigg(\frac{\rho'^2}{4}-\frac{\rho'\zeta'}{4}+\frac{\rho''}{2}-\frac{\zeta'}{r}+\frac{\rho'}{r}\bigg)
-P'_\bot\bigg(\frac{\zeta'}{4}-\frac{\rho'}{4}-\frac{3}{r}\bigg)+\frac{P''_\bot}{2}\bigg\}\bigg],
\end{align}
and for $\mathbb{L}_m=P_r$, we have
\begin{align}\nonumber
8\pi\mu&=e^{-\zeta}\bigg[\frac{\zeta'}{r}+\frac{e^\zeta}{r^2}-\frac{1}{r^2}+\varpi\bigg\{\mu\bigg(\frac{3\rho'\zeta'}{8}-\frac{\rho'^2}{8}
+\frac{\zeta'}{r}+\frac{e^\zeta}{r^2}-\frac{3\rho''}{4}-\frac{3\rho'}{2r}\\\nonumber
&-\frac{1}{r^2}\bigg)-\mu'\bigg(\frac{\zeta'}{4}-\frac{1}{r}-\rho'\bigg)+\frac{\mu''}{2}+P_r\bigg(\frac{\rho'\zeta'}{8}
-\frac{\rho'^2}{8}-\frac{\rho''}{4}-\frac{\zeta'}{2r}-\frac{e^\zeta}{r^2}\\\nonumber
&+\frac{1}{r^2}+\frac{\zeta''}{2}-\frac{3\zeta'^2}{4}\bigg)+\frac{5\zeta'P'_r}{4}-\frac{P''_r}{2}+P_\bot\bigg(\frac{\zeta'}{2r}-\frac{\rho'}{2r}
+\frac{3e^\zeta}{r^2}-\frac{1}{r^2}\bigg)\\\label{g9}
&-\frac{P'_\bot}{r}\bigg\}\bigg],\\\nonumber 8\pi
P_r&=e^{-\zeta}\bigg[\frac{\rho'}{r}-\frac{e^\zeta}{r^2}+\frac{1}{r^2}+\varpi\bigg\{\mu\bigg(\frac{\rho'\zeta'}{8}+\frac{\rho'^2}{8}
-\frac{\rho''}{4}-\frac{\rho'}{2r}\bigg)-\frac{\rho'\mu'}{4}-P_r\\\nonumber
&\times\bigg(\frac{5\rho'^2}{8}-\frac{7\rho'\zeta'}{8}+\frac{5\rho''}{4}-\frac{7\zeta'}{2r}+\frac{2\rho'}{r}-\zeta'^2
-\frac{2e^\zeta}{r^2}+\frac{2}{r^2}\bigg)+P'_r\bigg(\frac{\rho'}{4}+\frac{1}{r}\bigg)\\\label{g9a}
&-P_\bot\bigg(\frac{\zeta'}{2r}-\frac{\rho'}{2r}+\frac{3e^\zeta}{r^2}
-\frac{1}{r^2}\bigg)+\frac{P'_\bot}{r}\bigg\}\bigg],\\\nonumber 8\pi
P_\bot&=e^{-\zeta}\bigg[\frac{\rho'^2}{4}-\frac{\rho'\zeta'}{4}+\frac{\rho''}{2}-\frac{\zeta'}{2r}+\frac{\rho'}{2r}
+\varpi\bigg\{\mu\bigg(\frac{\rho'^2}{8}+\frac{\rho'\zeta'}{8}-\frac{\rho''}{4}-\frac{\rho'}{2r}\bigg)\\\nonumber
&-\frac{\rho'\mu'}{4}+P_r\bigg(\frac{\rho'\zeta'}{8}-\frac{\rho'^2}{8}-\frac{\rho''}{4}-\frac{\rho'}{2r}-\frac{\zeta''}{2}
+\frac{3\zeta'^2}{4}\bigg)-\frac{5\zeta'P'_r}{4}+\frac{P''_r}{2}\\\label{g9b}
&-P_\bot\bigg(\frac{\rho'^2}{4}-\frac{\rho'\zeta'}{4}+\frac{\rho''}{2}-\frac{\zeta'}{r}+\frac{\rho'}{r}\bigg)
-P'_\bot\bigg(\frac{\zeta'}{4}-\frac{\rho'}{4}-\frac{3}{r}\bigg)+\frac{P''_\bot}{2}\bigg\}\bigg].
\end{align}
The modified corrections involve state variables along with their
derivatives which make the above equations more complex. In
$f(\mathcal{R},\mathcal{T},\mathcal{Q})$ framework, Eq.\eqref{g7b}
produces the hydrostatic equilibrium condition corresponding to
$\mathbb{L}_m=-\mu$ as
\begin{align}\nonumber
&\frac{dP_r}{dr}+\frac{\rho'}{2}\left(\mu
+P_r\right)-\frac{2}{r}\left(P_\bot-P_r\right)-\frac{2\varpi
e^{-\zeta}}{\varpi\mathcal{R}+16\pi}\bigg[\frac{\rho'\mu}{8}\bigg(\rho'^2-\rho'\zeta'+2\rho''+\frac{4\rho'}{r}\bigg)\\\nonumber
&-\frac{\mu'}{8}\bigg(\rho'^2-\rho'\zeta'+2\rho''-\frac{4\rho'}{r}-\frac{8e^\zeta}{r^2}+\frac{8}{r^2}\bigg)+P_r\bigg(\frac{5\rho'^2\zeta'}{8}
-\frac{5\rho'\zeta'^2}{8}+\frac{7\rho''\zeta'}{4}\\\nonumber
&-\rho'\rho''+\frac{\rho'\zeta''}{2}-\frac{5\zeta'^2}{2r}-\frac{\rho'''}{2}+\frac{2\zeta''}{r}+\frac{\rho'\zeta'}{r}-\frac{\zeta'}{r^2}
-\frac{\rho''}{r}+\frac{\rho'}{r^2}+\frac{2e^\zeta}{r^3}-\frac{2}{r^3}\bigg)-\frac{P'_r}{8}\\\nonumber
&\times\bigg(\rho'^2-\rho'\zeta'+2\rho''-\frac{4\zeta'}{r}\bigg)+\frac{P_\bot}{r^2}\bigg(\zeta'-\rho'+\frac{2e^\zeta}{r}
-\frac{2}{r}\bigg)-\frac{P'_\bot}{r}\bigg(\frac{\zeta'}{2}-\frac{\rho'}{2}+\frac{e^\zeta}{r}\\\label{g11}
&-\frac{1}{r}\bigg)\bigg]=0,
\end{align}
and $\mathbb{L}_m=P_r$ provides as
\begin{align}\nonumber
&\frac{dP_r}{dr}+\frac{\rho'}{2}\left(\mu
+P_r\right)-\frac{2}{r}\left(P_\bot-P_r\right)-\frac{2\varpi
e^{-\zeta}}{\varpi\mathcal{R}+16\pi}\bigg[\frac{\rho'\mu}{8}\bigg(\rho'^2-\rho'\zeta'+2\rho''+\frac{4\rho'}{r}\bigg)\\\nonumber
&-\frac{\mu'}{8}\bigg(\rho'^2-\rho'\zeta'+2\rho''+\frac{4\rho'}{r}\bigg)+P_r\bigg(\frac{5\rho'^2\zeta'}{8}
-\frac{5\rho'\zeta'^2}{8}+\frac{7\rho''\zeta'}{4}+\frac{\rho'\zeta''}{2}-\frac{5\zeta'^2}{2r}\\\nonumber
&-\rho'\rho''-\frac{\rho'''}{2}+\frac{2\zeta''}{r}+\frac{\rho'\zeta'}{r}-\frac{\zeta'}{r^2}
-\frac{\rho''}{r}+\frac{\rho'}{r^2}+\frac{2e^\zeta}{r^3}-\frac{2}{r^3}\bigg)-\frac{P'_r}{8}\bigg(2\rho''-\rho'\zeta'\\\nonumber
&+\rho'^2-\frac{4\zeta'}{r}+\frac{8\rho'}{r}-\frac{8e^\zeta}{r^2}+\frac{8}{r^2}\bigg)+\frac{P_\bot}{r^2}\bigg(\zeta'-\rho'+\frac{2e^\zeta}{r}
-\frac{2}{r}\bigg)-\frac{P'_\bot}{r}\bigg(\frac{\zeta'}{2}-\frac{\rho'}{2}\\\label{g12}
&+\frac{e^\zeta}{r}-\frac{1}{r}\bigg)\bigg]=0.
\end{align}
Equations \eqref{g11} and \eqref{g12} are the generalized
Tolman-Opphenheimer-Volkoff ($\mathbb{TOV}$) equations in this
theory which can be utilized in studying the dynamics of
self-gravitating stars. The mass of spherical geometry provided by
Misner-Sharp \cite{41b} is
\begin{equation}\nonumber
m(r)=\frac{r}{2}\big(1-g^{\omega\alpha}r_{,\omega}r_{,\alpha}\big),
\end{equation}
which becomes
\begin{equation}\label{g12a}
m(r)=\frac{r}{2}\big(1-e^{-\zeta}\big).
\end{equation}

Different state variables associated with geometrical structures can
be interlinked through some relations, known as equations of state
which are significantly used to study the physical nature of compact
bodies. Among the resulting objects after death of massive star,
neutron stars are found as the most appealing structures in our
universe. They can be converted into black holes or quark stars
depending on their large or less densities, respectively
\cite{33b,41c}. Although these stars are surprisingly small, their
dense nature results in a strong gravitational field around them.
The non-linear systems of field equations \eqref{g8}-\eqref{g8b} and
\eqref{g9}-\eqref{g9b} encompass five unknown quantities such as
$\rho,~\zeta,~\mu,~P_r$ and $P_\bot$, therefore we need some
constraints to make the system definite. We assume that physical
variables in the interior geometry can be interlinked with the help
of $\mathbb{MIT}$ bag model $\mathbb{E}o\mathbb{S}$ and analyze the
properties of quark matter distribution \cite{33}. For this, the
quark pressure is given as
\begin{equation}\label{g13}
P_r=\sum_{\upsilon=u,d,s}P^\upsilon-\mathfrak{B_c},
\end{equation}
where $\mathfrak{B_c}$ symbolizes the bag constant. Further, the
quark matter is classified into three categories, namely up, down
and strange whose pressures are represented by $P^u,~P^d$ and $P^s$,
respectively. The density of each quark matter is linked with its
respective pressure by the relation as $\mu^l=3P^l$. Therefore, the
total density becomes
\begin{equation}\label{g14}
\mu=\sum_{\upsilon=u,d,s}\mu^\upsilon+\mathfrak{B_c}.
\end{equation}
Finally, we obtain the $\mathbb{MIT}$ bag model
$\mathbb{E}o\mathbb{S}$ for strange fluid by combining
Eqs.\eqref{g13} and \eqref{g14} as
\begin{equation}\label{g14a}
P_r=\frac{1}{3}\left(\mu-4\mathfrak{B_c}\right).
\end{equation}

Many researchers \cite{41f,41h} calculated the values of bag
constant for different stars and utilized them to analyze physical
features of these strange objects. We determine the exact solutions
to the field equations \eqref{g8}-\eqref{g8b} and
\eqref{g9}-\eqref{g9b} by using $\mathbb{E}o\mathbb{S}$
\eqref{g14a}. Equations \eqref{g8}-\eqref{g8b} thus produce the
solution as
\begin{align}\nonumber
\mu&=\bigg[8\pi
e^{\zeta}+\varpi\bigg(\frac{9\rho''}{8}-\frac{e^{\zeta}}{r^2}+\frac{1}{r^2}-\frac{\zeta''}{8}-\frac{5\rho'\zeta'}{8}-\frac{\zeta'^2}{16}
-\frac{7\zeta'}{2r}+\frac{3\rho'^2}{16}+\frac{7\rho'}{4r}\bigg)\bigg]^{-1}\\\nonumber
&\times\bigg[\frac{3}{4}\bigg(\frac{\zeta'}{r}+\frac{\rho'}{r}\bigg)+\mathfrak{B_c}\bigg\{8\pi
e^\zeta-\varpi\bigg(\frac{4\zeta'}{r}-\frac{3\rho'^2}{4}-\frac{3\rho''}{2}+\rho'\zeta'+\frac{\zeta''}{2}+\frac{\zeta'^2}{4}\\\label{g14b}
&-\frac{\rho'}{r}+\frac{e^\zeta}{r^2}-\frac{1}{r^2}\bigg)\bigg\}\bigg],\\\nonumber
P_r&=\bigg[8\pi
e^{\zeta}+\varpi\bigg(\frac{9\rho''}{8}-\frac{e^{\zeta}}{r^2}+\frac{1}{r^2}-\frac{\zeta''}{8}-\frac{5\rho'\zeta'}{8}-\frac{\zeta'^2}{16}
-\frac{7\zeta'}{2r}+\frac{3\rho'^2}{16}+\frac{7\rho'}{4r}\bigg)\bigg]^{-1}\\\label{g14c}
&\times\bigg[\frac{1}{4}\bigg(\frac{\zeta'}{r}+\frac{\rho'}{r}\bigg)-\mathfrak{B_c}\bigg\{8\pi
e^\zeta-\varpi\bigg(\frac{\rho'\zeta'}{2}
+\frac{\zeta'}{r}-\frac{2\rho'}{r}+\frac{e^\zeta}{r^2}-\rho''-\frac{1}{r^2}\bigg)\bigg\}\bigg],\\\nonumber
P_\bot&=\bigg[8\pi
e^{\zeta}+\varpi\bigg(\frac{\rho'^2}{4}-\frac{\rho'\zeta'}{4}+\frac{\rho''}{2}-\frac{\zeta'}{r}
+\frac{\rho'}{r}\bigg)\bigg]^{-1}\bigg[\frac{\rho'^2}{4}-\frac{\zeta'}{2r}+\frac{\rho'}{2r}-\frac{\rho'\zeta'}{4}+\frac{\rho''}{2}\\\nonumber
&+\varpi\bigg\{8\pi
e^{\zeta}+\varpi\bigg(\frac{9\rho''}{8}-\frac{e^{\zeta}}{r^2}+\frac{1}{r^2}-\frac{\zeta''}{8}-\frac{5\rho'\zeta'}{8}-\frac{\zeta'^2}{16}
-\frac{7\zeta'}{2r}+\frac{3\rho'^2}{16}+\frac{7\rho'}{4r}\bigg)\bigg\}^{-1}\\\nonumber
&\times\bigg\{\frac{1}{8r}\bigg(2\rho'\zeta'^2+\rho'^3-\rho''\zeta'-\rho'\rho''-\zeta'\zeta''-\rho'\zeta''
-\frac{3\rho'^2}{r}+\frac{3\zeta'^3}{2}+\frac{3\rho'^2\zeta'}{2}\\\nonumber
&-\frac{\zeta'^2}{r}-\frac{4\rho'\zeta'}{r}\bigg)+2\pi
e^\zeta\mathfrak{B_c}\bigg(\rho'\zeta'-2\rho''+2\zeta''-3\zeta'^2-\frac{2\rho'}{r}+\frac{2\zeta'}{r}\bigg)
+\frac{\varpi\mathfrak{B_c}}{16}\\\nonumber
&\times\bigg(10\rho''\zeta''-5\rho'\zeta'\zeta''+11\rho'\rho''\zeta'-11\rho''\zeta'^2-\rho'^2\zeta''
-2\rho''\rho'^2-10\rho''^2+\frac{\rho'^3\zeta'}{2}\\\nonumber
&-\frac{7\rho'^2\zeta'^2}{2}-\frac{36\rho'\zeta'^2}{r}-\frac{8\rho'^3}{r}+\frac{11\rho'\zeta'^3}{2}+\frac{16\rho'^2\zeta'}{r}
+\frac{28\rho''\zeta'}{r}-\frac{8\zeta'\zeta''}{r}-\frac{8\zeta''e^\zeta}{r^2}\\\nonumber
&+\frac{12\zeta'^3}{r}+\frac{3\rho'^4}{2}-\frac{8\rho'^2}{r^2}+\frac{8\zeta''}{r^2}-\frac{20\zeta'^2}{r^2}-\frac{24\rho'\rho''}{r}
+\frac{52\rho'\zeta'}{r^2}+\frac{10\rho'\zeta''}{r}+\frac{8e^\zeta\rho''}{r^2}\\\label{g14d}
&-\frac{4e^\zeta\rho'\zeta'}{r^2}-\frac{8\rho''}{r^2}+\frac{12\zeta'^2e^\zeta}{r^2}-\frac{8\rho'}{r^3}
-\frac{8e^\zeta\zeta'}{r^3}+\frac{8\zeta'}{r^3}+\frac{8e^\zeta\rho'}{r^3}\bigg)\bigg\}\bigg],
\end{align}
and Eqs.\eqref{g9}-\eqref{g9b} yield
\begin{align}\nonumber
\mu&=\bigg[8\pi
e^{\zeta}+\varpi\bigg(\frac{9\rho''}{8}-\frac{e^{\zeta}}{r^2}+\frac{1}{r^2}-\frac{\zeta''}{8}-\frac{5\rho'\zeta'}{8}-\frac{\zeta'^2}{16}
-\frac{3\zeta'}{2r}+\frac{3\rho'^2}{16}+\frac{2\rho'}{r}\bigg)\bigg]^{-1}\\\nonumber
&\times\bigg[\frac{3}{4}\bigg(\frac{\zeta'}{r}+\frac{\rho'}{r}\bigg)+\mathfrak{B_c}\bigg\{8\pi
e^\zeta-\varpi\bigg(\frac{3\zeta'}{r}-\frac{3\rho'^2}{4}-\frac{3\rho''}{2}+\rho'\zeta'+\frac{\zeta''}{2}+\frac{\zeta'^2}{4}\\\label{g14e}
&-\frac{2\rho'}{r}+\frac{e^\zeta}{r^2}-\frac{1}{r^2}\bigg)\bigg\}\bigg],\\\nonumber
P_r&=\bigg[8\pi
e^{\zeta}+\varpi\bigg(\frac{9\rho''}{8}-\frac{e^{\zeta}}{r^2}+\frac{1}{r^2}-\frac{\zeta''}{8}-\frac{5\rho'\zeta'}{8}-\frac{\zeta'^2}{16}
-\frac{3\zeta'}{2r}+\frac{3\rho'^2}{16}+\frac{2\rho'}{r}\bigg)\bigg]^{-1}\\\label{g14f}
&\times\bigg[\frac{1}{4}\bigg(\frac{\zeta'}{r}+\frac{\rho'}{r}\bigg)-\mathfrak{B_c}\bigg\{8\pi
e^\zeta+\varpi\bigg(\rho''-\frac{\rho'\zeta'}{2}
-\frac{\zeta'}{r}+\frac{2\rho'}{r}-\frac{e^\zeta}{r^2}+\frac{1}{r^2}\bigg)\bigg\}\bigg],\\\nonumber
P_\bot&=\bigg[8\pi
e^{\zeta}+\varpi\bigg(\frac{\rho'^2}{2}-\frac{\rho'\zeta'}{4}+\frac{\rho''}{2}-\frac{\zeta'}{r}
+\frac{\rho'}{r}\bigg)\bigg]^{-1}\bigg[\frac{\rho'^2}{4}-\frac{\zeta'}{2r}+\frac{\rho'}{2r}-\frac{\rho'\zeta'}{4}+\frac{\rho''}{2}\\\nonumber
&+\varpi\bigg\{8\pi
e^{\zeta}+\varpi\bigg(\frac{9\rho''}{8}-\frac{e^{\zeta}}{r^2}+\frac{1}{r^2}-\frac{\zeta''}{8}-\frac{5\rho'\zeta'}{8}-\frac{\zeta'^2}{16}
-\frac{3\zeta'}{2r}+\frac{3\rho'^2}{16}+\frac{2\rho'}{r}\bigg)\bigg\}^{-1}\\\nonumber
&\times\bigg\{\frac{1}{8r}\bigg(\frac{5\rho'\zeta'^2}{2}+\frac{\rho'^3}{2}-2\rho''\zeta'-2\rho'\rho''-\zeta'\zeta''-\rho'\zeta''
-\frac{4\rho'^2}{r}+\frac{3\zeta'^3}{2}+\frac{3\rho'^2\zeta'}{2}\\\nonumber
&-\frac{4\rho'\zeta'}{r}\bigg)+2\pi
e^\zeta\mathfrak{B_c}\bigg(\rho'^2+2\zeta''-3\zeta'^2\bigg)+\frac{\varpi\mathfrak{B_c}}{16}\bigg\{\bigg(\rho'\zeta'+\frac{3\zeta'}{r}
-\frac{3\rho'^2}{4}+\frac{e^\zeta}{r^2}\\\nonumber
&-\frac{1}{r^2}-\frac{3\rho''}{2}+\frac{\zeta''}{2}+\frac{\zeta'^2}{4}-\frac{2\rho'}{r}\bigg)\bigg(\frac{\rho''}{4}-\frac{\rho'\zeta'}{8}
-\frac{\rho'^2}{8}+\frac{\rho'}{2r}\bigg)+\bigg(\frac{\zeta'}{r}+\frac{e^\zeta}{r^2}-\frac{1}{r^2}\\\label{g14g}
&+\frac{\rho'\zeta'}{2}-\frac{2\rho'}{r}-\rho''\bigg)\bigg(\frac{\rho'\zeta'}{8}-\frac{\rho'^2}{8}-\frac{\rho''}{4}-\frac{\rho'}{2r}
-\frac{\zeta''}{2}+\frac{3\zeta'^2}{4}\bigg)\bigg\}\bigg\}\bigg].
\end{align}
Various researchers have frequently used the $\mathbb{E}o\mathbb{S}$
\eqref{g14a} in $\mathbb{GR}$ and modified scenarios such as
$f(\mathcal{R}),~f(\mathcal{G})$ and $f(\mathcal{R},\mathcal{T})$
theories to examine the matter configuration inside the quark
bodies. We utilize this $\mathbb{E}o\mathbb{S}$ to develop solutions
to the field equations for the considered choices of $\mathbb{L}_m$
and check their physical feasibility for both values of the coupling
constant.

\subsection{Embedding Class-one Condition}

If the Gauss-Codazzi equations (also called the
Gauss-Codazzi-Mainardi equations through which the induced metric
and second fundamental form of a submanifold can be linked together)
\begin{equation}
\mathcal{R}_{\rho\zeta\alpha\beta}=2\mathbf{e}\mathcal{Q}_{\rho[\alpha}\mathcal{Q}_{\beta]\zeta},
\quad
\mathcal{Q}_{\rho[\zeta;\alpha]}-\Gamma^\beta_{\zeta\alpha}\mathcal{Q}_{\rho\beta}+\Gamma^\beta_{\rho[\zeta}\mathcal{Q}_{\alpha]\beta}=0,
\end{equation}
are satisfied by a symmetric tensor $\mathcal{Q}_{\rho\zeta}$, then
($n-2$)-dimensional space can be embedded into an
$(n-1)$-dimensional space. Here,
$\mathcal{R}_{\rho\zeta\alpha\beta}$ shows curvature tensor,
$\mathcal{Q}_{\rho\zeta}$ denotes the coefficients of second
differential form and $\mathbf{e}=\pm1$. The first equation, often
called the Gauss equation, says that the derivatives of the Gauss
map at any given point determines the Gauss curvature of the surface
at that point. The second equation, called the Codazzi (or
Codazzi-Mainardi) equation, states that the covariant derivative of
the second fundamental form is fully symmetric.

The necessary and sufficient condition for an embedding class-one is
computed by Eiesland \cite{41i} as
\begin{equation}
R_{0101}R_{2323}-R_{1212}R_{0303}-R_{1202}R_{1303},
\end{equation}
which produces the differential equation in terms of metric
coefficients ($\rho,\zeta$) as
\begin{equation}
\big(\zeta'-\rho'\big)\rho'e^\zeta+2\big(1-e^\zeta\big)\rho''+\rho'^2=0.
\end{equation}
The above equation provides a solution as
\begin{equation}\label{g14h}
\zeta(r)=\ln\big(1+X\rho'^2e^\rho\big),
\end{equation}
where $X$ is an integration constant. We consider one of the metric
functions proposed by Maurya \emph{et al.} \cite{37,37a} as
\begin{equation}\label{g14i}
\rho(r)=2Wr^2+\ln Y,
\end{equation}
where $W$ and $Y$ are positive unknowns. To check the acceptability
criteria (suggested by Lake \cite{41j}) of the considered form of
$\rho(r)$, we take its differentials as
\begin{equation}\nonumber
\rho'(r)=4Wr, \quad \rho''(r)=4W,
\end{equation}
from which we observe that $\rho(0)=\ln Y,~\rho'(0)=0$ and
$\rho''(0)>0$ within the whole configuration, where $r=0$ is center
of the star. Hence, the metric potential \eqref{g14i} is acceptable.
By combining Eqs.\eqref{g14h} and \eqref{g14i}, we obtain $\zeta(r)$
as
\begin{equation}\label{g15}
\zeta(r)=\ln\big(1+WZr^2e^{2Wr^2}\big),
\end{equation}
where $Z=16WXY$. The field equations \eqref{g14b}-\eqref{g14d} and
\eqref{g14e}-\eqref{g14g} in terms of metric functions \eqref{g14i}
and \eqref{g15} are given in Appendix \textbf{A}.

\section{Boundary Conditions}

The formation of anisotropic configured astronomical structures can
be understood in a better way by matching their inner and outer
geometries smoothly. It is assumed that the spacetime outside the
geometry \eqref{g6} is empty, thus we take Schwarzschild metric
which is defined as
\begin{equation}\label{g20}
ds^2=-\left(1-\frac{2\bar{M}}{r}\right)dt^2+\frac{dr^2}{\left(1-\frac{2\bar{M}}{r}\right)^{-1}}
+r^2d\theta^2+r^2\sin^2\theta d\psi^2,
\end{equation}
where $\bar{M}(r)$ shows the total mass of the considered geometry
at boundary ($r=\mathcal{H}$). As the metric components of
geometries \eqref{g6} and \eqref{g20} are continuous across the
boundary, thus we obtain the following constraints
\begin{eqnarray}\label{g21}
g_{tt}&=&e^{\rho(r)}=Ye^{2W\mathcal{H}^2}=1-\frac{2\bar{M}}{\mathcal{H}},\\\label{g21a}
g_{rr}&=&e^{\zeta(r)}=1+WZ\mathcal{H}^2e^{2W\mathcal{H}^2}=\bigg(1-\frac{2\bar{M}}{\mathcal{H}}\bigg)^{-1},\\\label{g22}
\frac{\partial g_{tt}}{\partial
r}&=&\rho'(r)=4W\mathcal{H}=\frac{2\bar{M}}{\mathcal{H}\big(\mathcal{H}-2\bar{M}\big)}.
\end{eqnarray}
We determine the four unknowns ($W,X,Y,Z$) by solving
Eqs.\eqref{g21}-\eqref{g22} simultaneously as
\begin{eqnarray}\label{g23}
W&=&\frac{\bar{M}}{2\mathcal{H}^2\big(\mathcal{H}-2\bar{M}\big)},\\\label{g24}
X&=&\frac{\mathcal{H}^3}{2\bar{M}},\\\label{g25}
Y&=&\bigg(\frac{\mathcal{H}-2\bar{M}}{\mathcal{H}}\bigg)e^{\frac{\bar{M}}{2\bar{M}-\mathcal{H}}},\\\label{g25a}
Z&=&4e^{\frac{\bar{M}}{2\bar{M}-\mathcal{H}}}.
\end{eqnarray}
The radial pressure in the interior of compact stars must vanish at
the boundary ($r=\mathcal{H}$), thus we obtain the value of bag
constant from Eq.\eqref{g14c} along with
Eqs.\eqref{g23}-\eqref{g25a} as
\begin{align}\label{g26}
\mathfrak{B_c}=-\frac{\bar{M}(2\bar{M}-3\mathcal{H})(2\bar{M}-\mathcal{H})}{4\left(6\varpi\bar{M}^3-8\varpi\bar{M}^2\mathcal{H}
+8\pi\bar{M}\mathcal{H}^4+3\varpi\bar{M}\mathcal{H}^2-4\pi\mathcal{H}^5\right)},
\end{align}
while Eq.\eqref{g14f} provides its value as
\begin{align}\label{g27}
\mathfrak{B_c}=-\frac{\bar{M}(2\bar{M}-3\mathcal{H})(2\bar{M}-\mathcal{H})}{8\left(\varpi\bar{M}^3+4\pi\bar{M}\mathcal{H}^4
-2\pi\mathcal{H}^5\right)}.
\end{align}

We utilize the experimental data of each strange star such as their
masses and radii to calculate the values of $W,~X,~Y,~Z$ and
$\mathfrak{B_c}$, as presented in Table $\mathbf{1}$. It is noticed
that all the compact bodies show compatible behavior with the limit
proposed by Buchdhal \cite{42a}, i.e.,
$\frac{2\bar{M}}{\mathcal{H}}<\frac{8}{9}$. We obtain two solutions
and evaluate their corresponding values of $\mathfrak{B_c}$ by using
Eqs.\eqref{g26} and \eqref{g27} to analyze the stellar evolution.
Table $\mathbf{2}$ presents the corresponding values of four
constants involving in embedding condition. The bag constant, energy
density at the center as well as surface and radial pressure at the
center for each star candidate corresponding to $\mathbb{L}_m=-\mu$
and $P_r$ are provided in Tables $\mathbf{3}$ and $\mathbf{4}$,
respectively. For massless quarks, the bag constant has the range
58.9-91.5 $MeV/fm^3$ \cite{42aa}, whereas it lies within $56-78$
$MeV/fm^3$ for massive quarks (with approximate mass as $150$ $MeV$)
\cite{42ab}. For $\varpi=4$, we have values of the bag constant as
follows
\begin{itemize}
\item For $\mathbb{L}_m=-\mu$, these values are $116.73,~63.49,~217.12,~238.11$ and $113.83$
$MeV/fm^3$, respectively.
\item For $\mathbb{L}_m=P_r$, these values are $116.37,~63.38,~215.88,~236.61$ and $113.49$
$MeV/fm^3$, respectively.
\end{itemize}
It can be identified that their observed values for which stars stay
stable are much smaller than the above calculated values, except for
the candidate Cen X-3. We can also see that there is a little
difference between the values of bag constant for both choices of
the matter Lagrangian. Nonetheless, some experiments have been
performed by $\mathrm{CERN-SPS}$ and $\mathrm{RHIC}$, and it was
concluded that the density dependent bag model may offer a broad
range of values of the bag constant.
\begin{table}[H]
\scriptsize \centering \caption{Physical values of different compact
star candidates} \label{Table1} \vspace{+0.1in}
\setlength{\tabcolsep}{0.95em}
\begin{tabular}{cccccc}
\hline\hline Star Models & 4U 1820-30 & Cen X-3 & SAX J 1808.4-3658
& RXJ 1856-37 & Her X-I
\\\hline $Mass(M_{\bigodot})$ & 2.25 & 1.49 & 1.435 & 0.9041 & 0.88
\\\hline
$\mathcal{H}(km)$ & 10 & 11.06 & 7.07 & 6 & 7.7
\\\hline
$\bar{M}/\mathcal{H}$ & 0.331 & 0.198 & 0.298 & 0.222 & 0.168  \\
\hline\hline
\end{tabular}
\end{table}
\begin{table}[H]
\scriptsize \centering \caption{Calculated values of constants
$W,~X,~Y$ and $Z$ for different compact star candidates}
\label{Table2} \vspace{+0.1in} \setlength{\tabcolsep}{0.95em}
\begin{tabular}{cccccc}
\hline\hline Star Models & 4U 1820-30 & Cen X-3 & SAX J 1808.4-3658
& RXJ 1856-37 & Her X-I
\\\hline $W$ & 0.0048855 & 0.0013404 & 0.00740096 & 0.00552334 &
0.00213368
\\\hline
$X$ & 151.172 & 308.839 & 83.764 & 81.2623 & 176.458
\\\hline
$Y$ & 0.127411 & 0.435080 & 0.192429 & 0.374229 & 0.515568
\\\hline
$Z$ & 1.5056 & 2.8817 & 1.9087 & 2.6875 & 3.1058  \\
\hline\hline
\end{tabular}
\end{table}
\begin{table}[H]
\scriptsize \centering \caption{Physical parameters and bag constant
of different compact star candidates corresponding to
$\mathbb{L}_m=-\mu$ and $\varpi=4$} \label{Table3} \vspace{+0.1in}
\setlength{\tabcolsep}{0.95em}
\begin{tabular}{cccccc}
\hline\hline Star Models & 4U 1820-30 & Cen X-3 & SAX J 1808.4-3658
& RXJ 1856-37 & Her X-I
\\\hline $\mathfrak{B_c}$ & 0.000154482 & 0.000084016 & 0.000287334 & 0.000315105 &
0.000150640
\\\hline
$\mu_c (gm/cm^3)$ & 1.5733$\times$10$^{15}$ &
6.2249$\times$10$^{14}$ & 2.6957$\times$10$^{15}$ &
2.4964$\times$10$^{15}$ & 1.0768$\times$10$^{15}$
\\\hline
$\mu_s (gm/cm^3)$ & 8.2129$\times$10$^{14}$ &
4.8042$\times$10$^{14}$ & 7.7862$\times$10$^{14}$ &
8.3548$\times$10$^{14}$ & 6.5086$\times$10$^{14}$
\\\hline
$P_{rc} (dyne/cm^2)$ & 2.2221$\times$10$^{35}$ &
5.4133$\times$10$^{34}$ & 3.4774$\times$10$^{35}$ &
2.4433$\times$10$^{35}$ & 8.0935$\times$10$^{34}$
\\\hline
$\sigma_s$ & 0.0145 & 0.0068 & 0.0177 & 0.0167 & 0.0102
\\\hline
$D_s$ & 0.0154 & 0.0071 & 0.0189 & 0.0178 & 0.0107  \\
\hline\hline
\end{tabular}
\end{table}
\begin{table}[H]
\scriptsize \centering \caption{Physical parameters and bag constant
of different compact star candidates corresponding to
$\mathbb{L}_m=P_r$ and $\varpi=4$} \label{Table4} \vspace{+0.1in}
\setlength{\tabcolsep}{0.95em}
\begin{tabular}{cccccc}
\hline\hline Star Models & 4U 1820-30 & Cen X-3 & SAX J 1808.4-3658
& RXJ 1856-37 & Her X-I
\\\hline $\mathfrak{B_c}$ & 0.000154006 & 0.000083875 & 0.000285693 & 0.000313131 & 0.000150188
\\\hline
$\mu_c (gm/cm^3)$ & 1.3914$\times$10$^{15}$ &
5.5226$\times$10$^{14}$ & 2.3787$\times$10$^{15}$ &
2.2021$\times$10$^{15}$ & 9.4452$\times$10$^{14}$
\\\hline
$\mu_s (gm/cm^3)$ & 7.2805$\times$10$^{14}$ &
4.1701$\times$10$^{14}$ & 6.8029$\times$10$^{14}$ &
7.2805$\times$10$^{14}$ & 5.7928$\times$10$^{14}$
\\\hline
$P_{rc} (dyne/cm^2)$ & 2.1114$\times$10$^{35}$ &
5.2281$\times$10$^{34}$ & 3.2754$\times$10$^{35}$ &
2.2954$\times$10$^{35}$ & 7.7231$\times$10$^{34}$
\\\hline
$\sigma_s$ & 0.0129 & 0.0061 & 0.0156 & 0.0147 & 0.0089
\\\hline
$D_s$ & 0.0139 & 0.0064 & 0.0171 & 0.0159 & 0.0096  \\
\hline\hline
\end{tabular}
\end{table}

\section{Physical Analysis of Compact Stars}
\begin{figure}\center
\epsfig{file=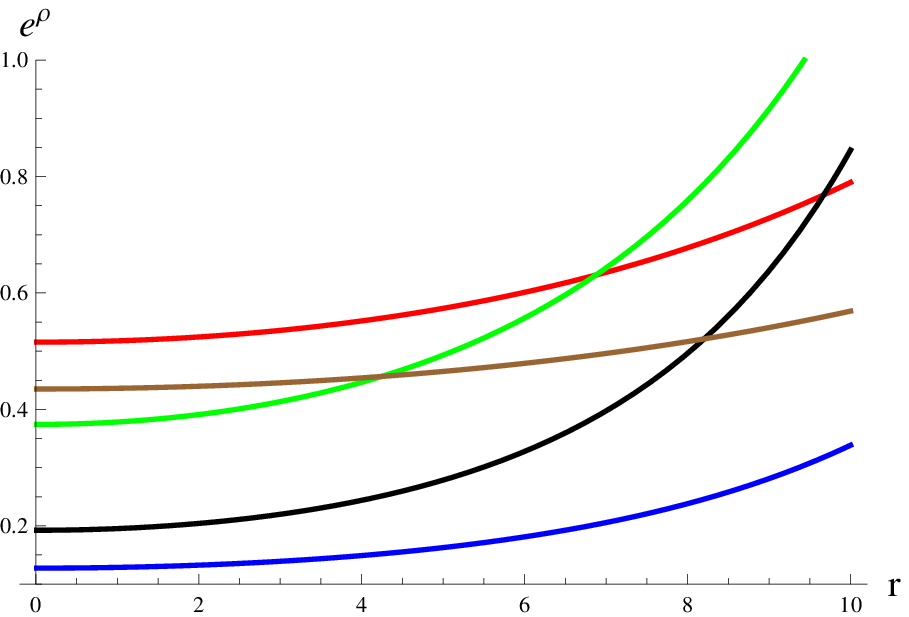,width=0.45\linewidth}\epsfig{file=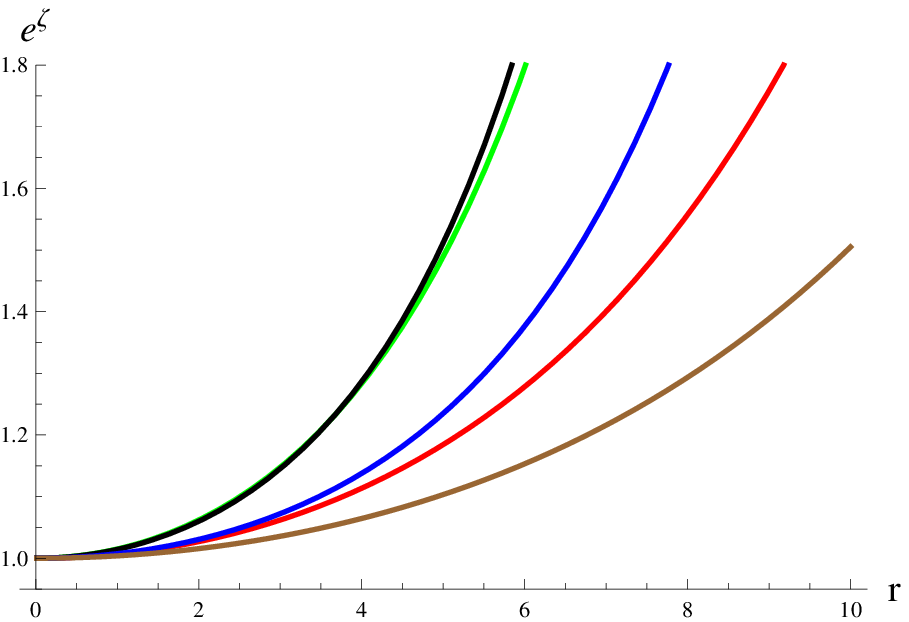,width=0.45\linewidth}
\caption{Metric potentials versus $r$ for different compact star
candidates}
\end{figure}

This section is related to the study of various physical
characteristics of the considered compact stars associated with
anisotropic distribution in their interiors in the framework of
$f(\mathcal{R},\mathcal{T},\mathcal{R}_{\omega\alpha}\mathcal{T}^{\omega\alpha})$
theory. We analyze the graphical behavior of both developed
solutions \eqref{g14b}-\eqref{g14d} and \eqref{g14e}-\eqref{g14g}
for $\varpi=\pm4$ corresponding to all stars by using their
respective preliminary data provided in Tables $\mathbf{1}$ and
$\mathbf{2}$. Further, we check the physical behavior of temporal as
well as radial metric functions, anisotropy, energy conditions and
mass in the interior of all quark candidates. For particular values
of the model parameter, we also analyze the stability of resulting
solutions. It is familiar that the resulting solution will be
assumed compatible if the metric potential possesses non-singular
and increasing nature in the whole positive domain. In this case,
the metric coefficients are presented in Eqs.\eqref{g14i} and
\eqref{g15} involving four constants which are calculated in Table
$\mathbf{2}$. Figure $\mathbf{1}$ exhibits their plots from which we
observe that our resulting solutions are physically consistent. It
is worth mentioning here that the brown color expresses Cen X-3
compact star, blue indicates 4U 1820-30, red signifies Her X-I,
black represents SAX J 1808.4-3658 and green color shows RXJ 1856-37
in all plots.
\begin{figure}\center
\epsfig{file=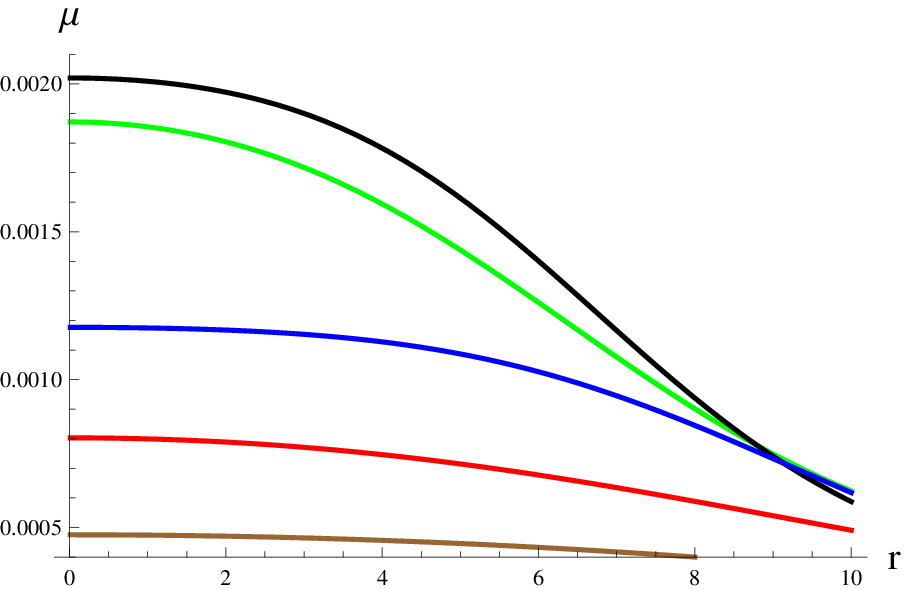,width=0.45\linewidth}\epsfig{file=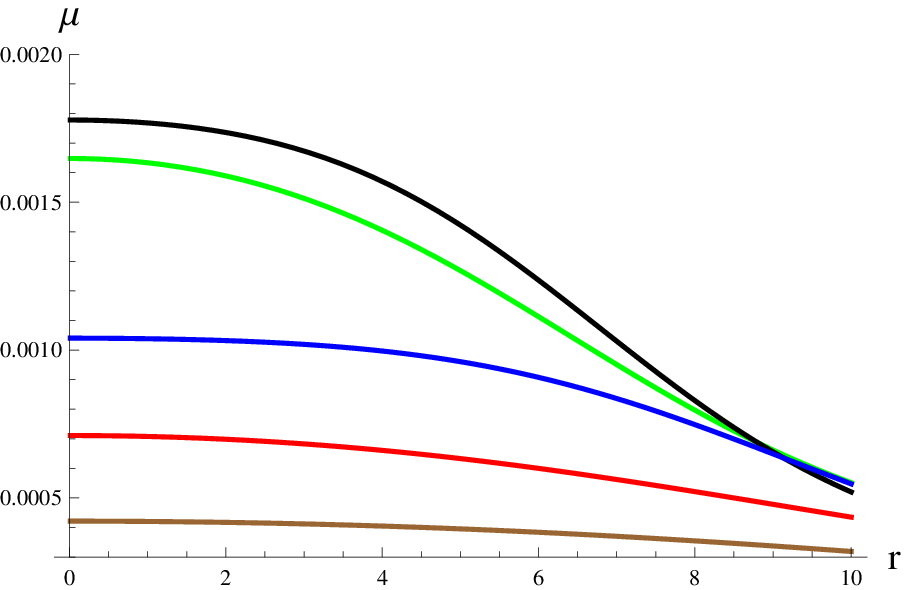,width=0.45\linewidth}
\epsfig{file=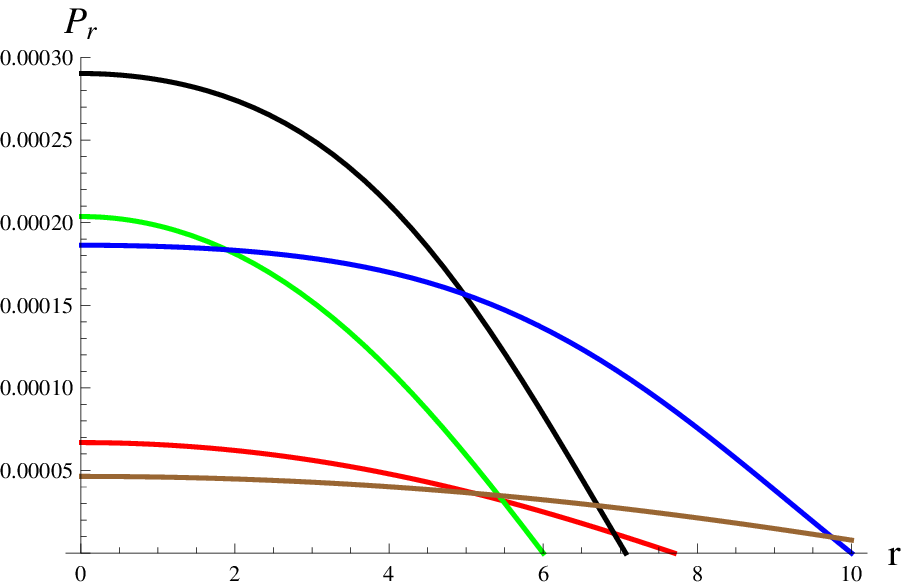,width=0.45\linewidth}\epsfig{file=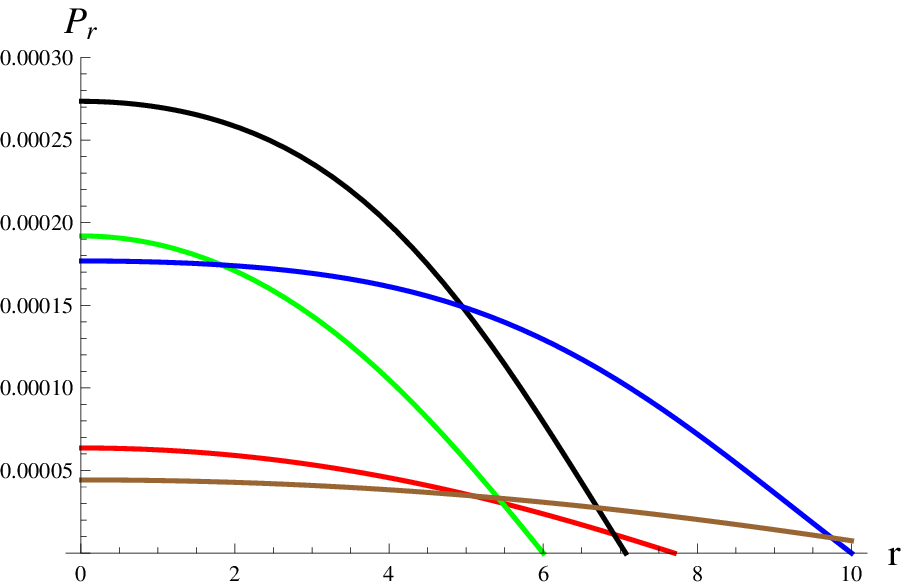,width=0.45\linewidth}
\epsfig{file=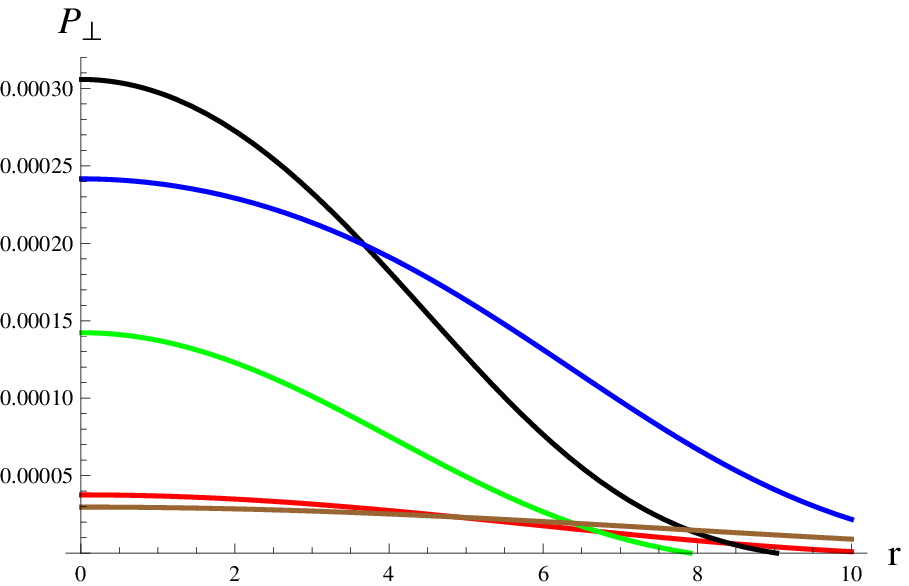,width=0.45\linewidth}\epsfig{file=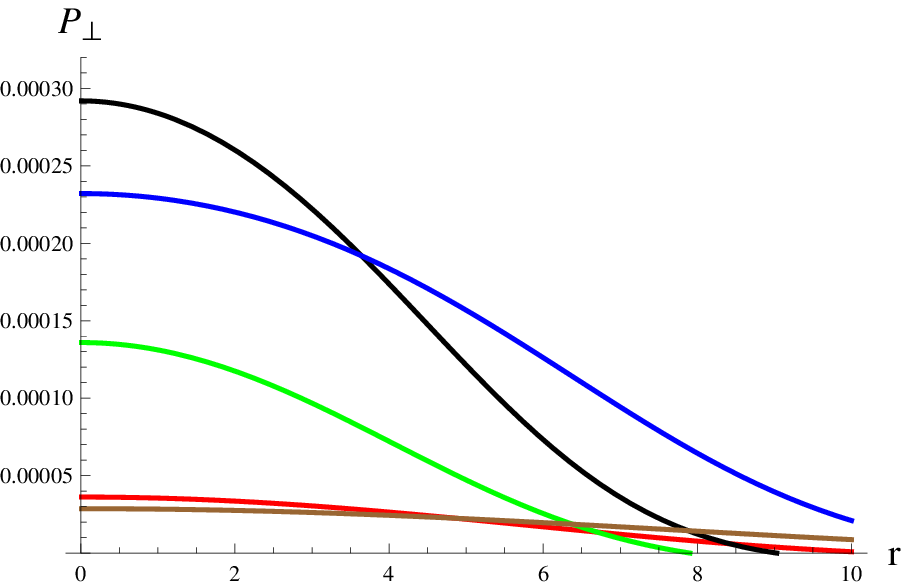,width=0.45\linewidth}
\caption{Plots of energy density, radial pressure and tangential
pressure versus $r$ corresponding to $\varpi=4$ and
$\mathbb{L}_m=-\mu$ (left) as well as $\mathbb{L}_m=P_r$ (right) for
different compact star candidates}
\end{figure}

\subsection{Study of Physical Variables and their Regularity Conditions}

The physically acceptable solution guarantees the maximum value of
matter variables like energy density and pressure at the center and
minimum value at the boundary of self-gravitating stellar
structures. Figure $\mathbf{2}$ assures that $\varpi=4$ provides
acceptable solutions corresponding to each star for both choices of
matter Lagrangian, as all variables fulfill the above acceptability
criteria, thus these compact stars have extremely dense structures
in modified gravity. Figure $\mathbf{2}$ (two plots in second row)
also show the disappearance of radial pressure at the surface of
each candidate. In Tables $\mathbf{3}$ and $\mathbf{4}$, the
calculated values of $\mu_c,~\mu_s$ and $P_{rc}$ are provided with
respect to both solutions, which indicate the energy density at
center, at surface and radial pressure at center, respectively. We
can see from these tables that the solution corresponding to
$\mathbb{L}_m=-\mu$ provides more dense structure of each star. To
show regular behavior of the solution, some conditions at the center
should be satisfied as $\frac{d\mu}{dr}|_{r=0} =
0,~\frac{dP_r}{dr}|_{r=0} = 0,~\frac{d^2\mu}{dr^2}|_{r=0} < 0$ and
$\frac{d^2P_r}{dr^2}|_{r=0} < 0$. Figures $\mathbf{3}$ and
$\mathbf{4}$ reveal that both solutions fulfill maximality
conditions.

The graphical nature of the matter variables and their differentials
corresponding to both the obtained solutions is also checked for
$\varpi=-4$ and found to be acceptable, but their graphs are not
added in this article.
\begin{figure}\center
\epsfig{file=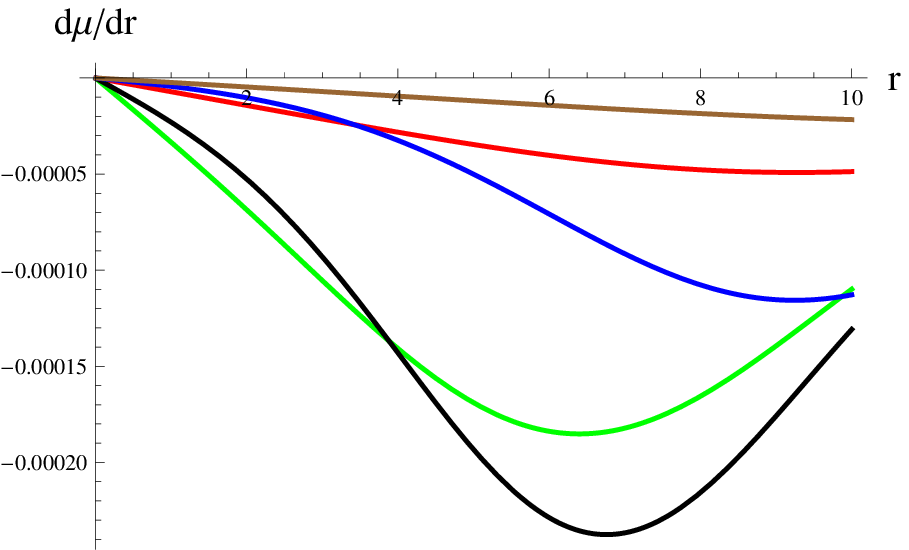,width=0.45\linewidth}\epsfig{file=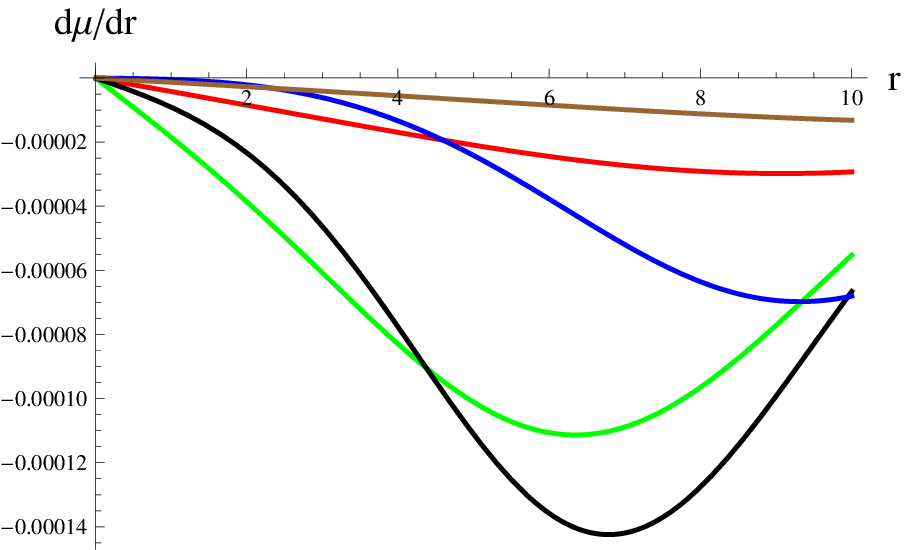,width=0.45\linewidth}
\epsfig{file=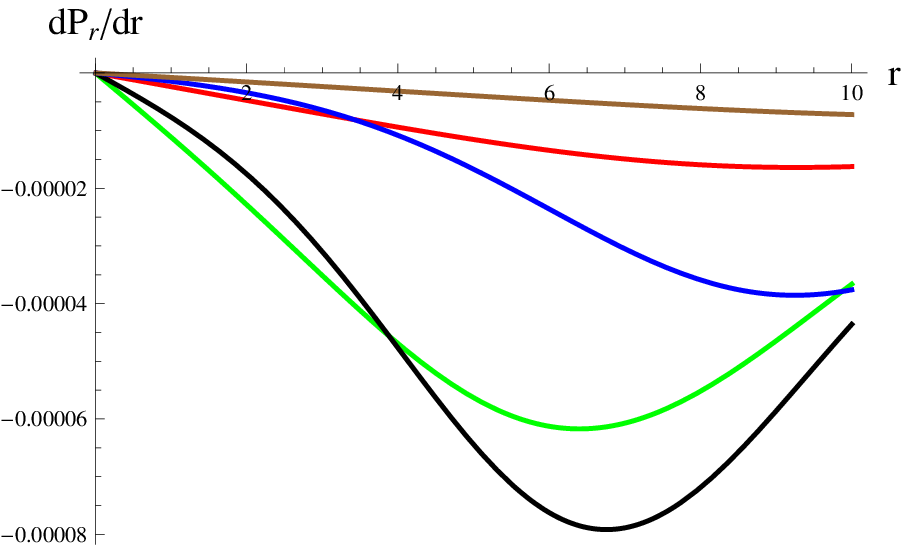,width=0.45\linewidth}\epsfig{file=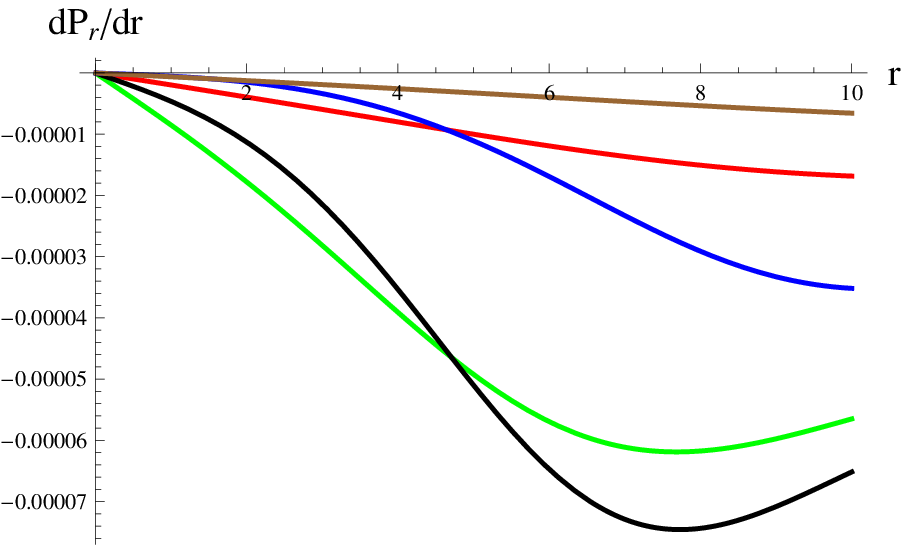,width=0.45\linewidth}
\epsfig{file=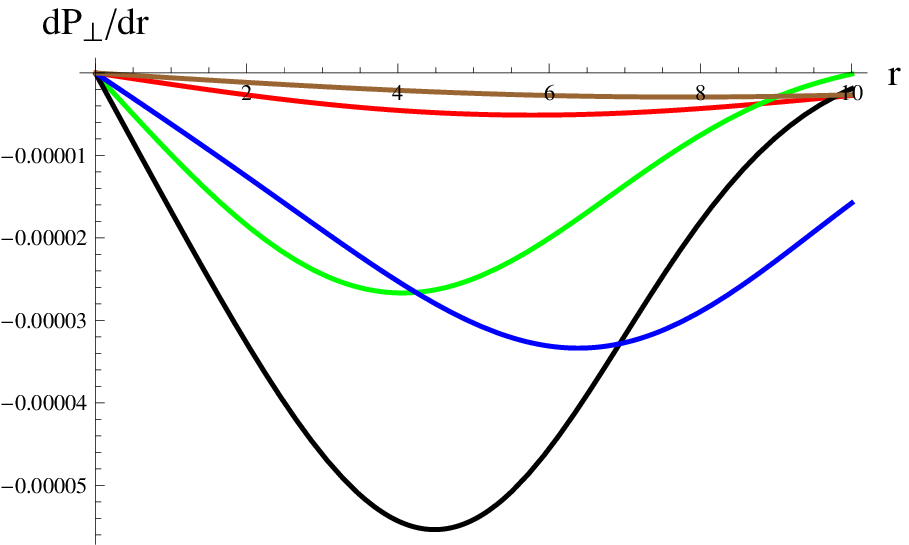,width=0.45\linewidth}\epsfig{file=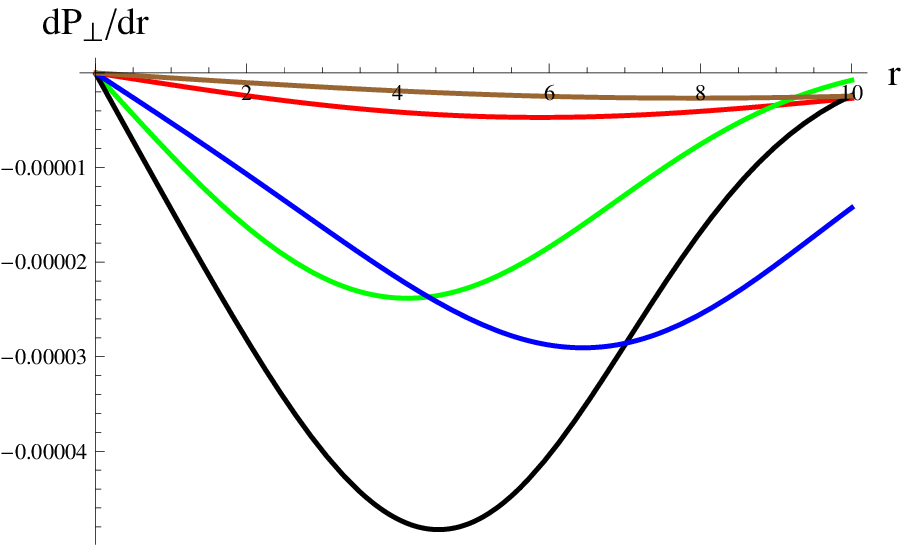,width=0.45\linewidth}
\caption{Plots of $\frac{d\mu}{dr},~\frac{dP_r}{dr}$ and
$\frac{dP_\bot}{dr}$ versus $r$ corresponding to $\varpi=4$ and
$\mathbb{L}_m=-\mu$ (left) as well as $\mathbb{L}_m=P_r$ (right) for
different compact star candidates}
\end{figure}
\begin{figure}\center
\epsfig{file=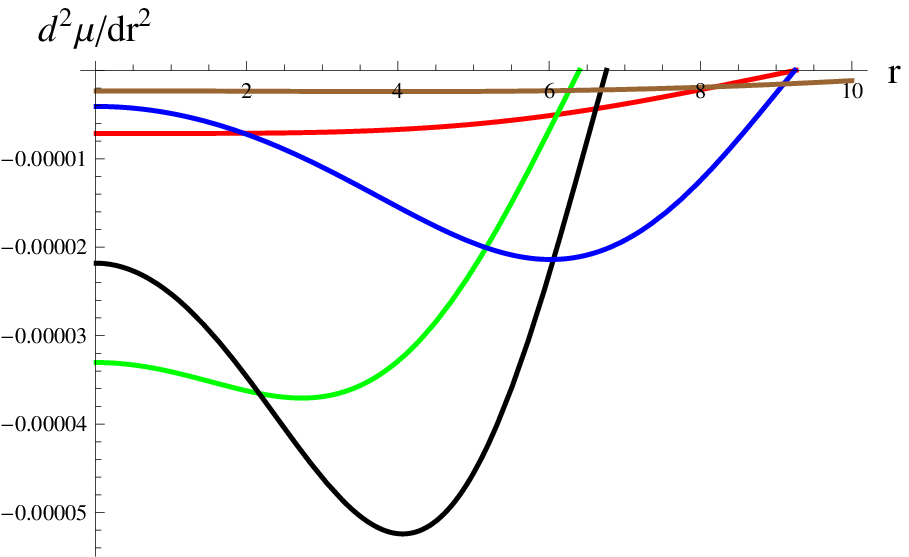,width=0.45\linewidth}\epsfig{file=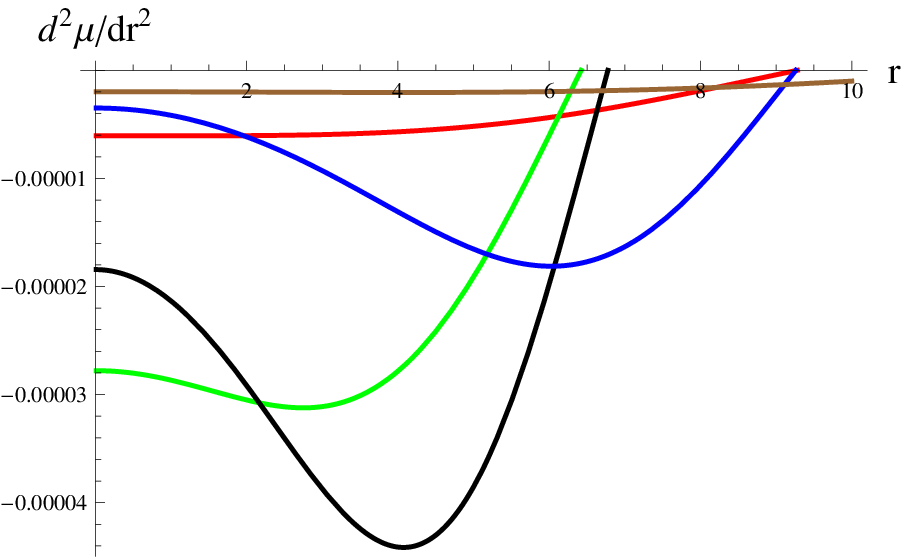,width=0.45\linewidth}
\epsfig{file=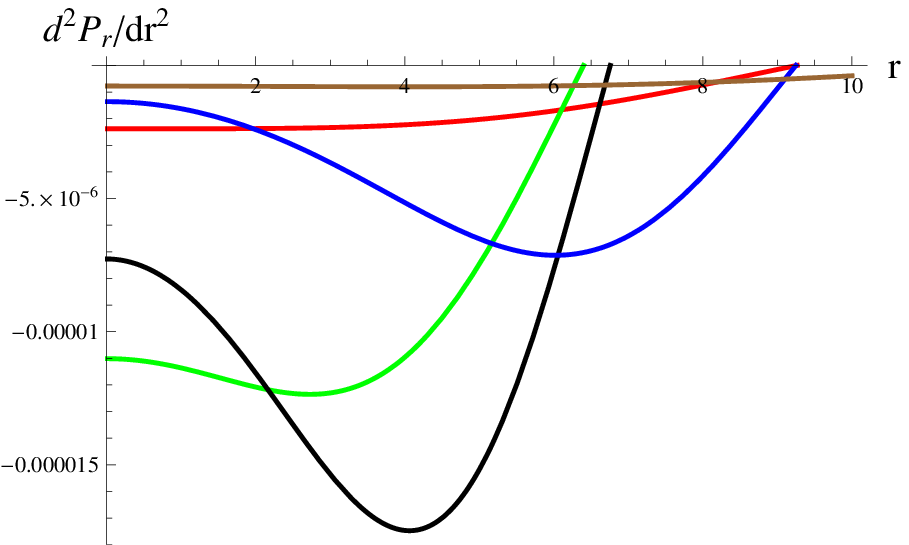,width=0.45\linewidth}\epsfig{file=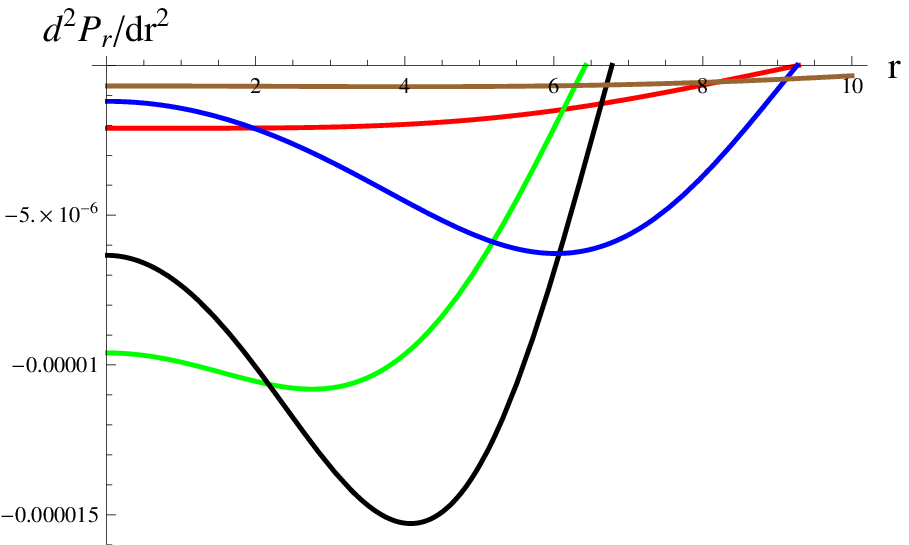,width=0.45\linewidth}
\epsfig{file=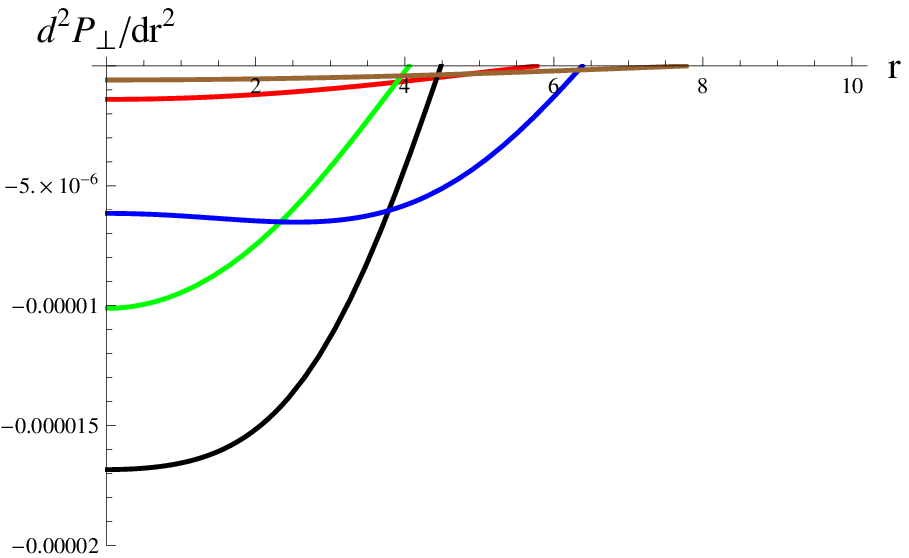,width=0.45\linewidth}\epsfig{file=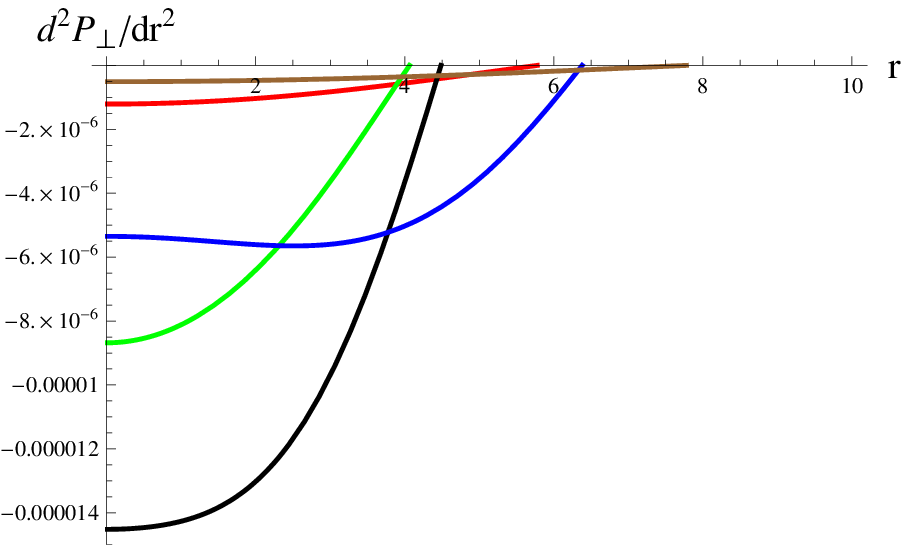,width=0.45\linewidth}
\caption{Plots of $\frac{d^2\mu}{dr^2},~\frac{d^2P_r}{dr^2}$ and
$\frac{d^2P_\bot}{dr^2}$ versus $r$ corresponding to $\varpi=4$ and
$\mathbb{L}_m=-\mu$ (left) as well as $\mathbb{L}_m=P_r$ (right) for
different compact star candidates}
\end{figure}

\subsection{Effect of Anisotropic Pressure}

For the first solution corresponding to $\mathbb{L}_m=-\mu$, the
anisotropy (i.e., $\Delta=P_\bot-P_r$) is obtained as
\begin{align}\nonumber
\Delta&=\bigg[2\big(\varpi
W\big(-Ze^{2r^2W}+2r^2W+3\big)+4\pi\big(r^2WZe^{2r^2W}+1\big)^2\big)\big(32\pi\big(r^2WZ\\\nonumber
&\times e^{2r^2W}+1\big)^3-\varpi
W\big(2r^2W\big(14Z^2e^{4r^2W}+Ze^{2r^2W}-6\big)+25Ze^{2r^2W}+32\\\nonumber
&\times
r^6W^3Z^2e^{4r^2W}+4r^4W^2Ze^{2r^2W}\big(Z^2e^{4r^2W}+8Ze^{2r^2W}+6\big)-34\big)\big)\bigg]^{-1}\bigg[256\\\nonumber
&\times \pi^2\mathfrak{B_c}\big(r^2WZe^{2r^2W}+1\big)^5-16\pi
W\big(r^2WZe^{2r^2W}+1\big)^2\big(-28\varpi\mathfrak{B_c}+r^2W\\\nonumber
&\times\big(Ze^{2r^2W}-2\big)\big(8\varpi\mathfrak{B_c}+3Z(8\varpi\mathfrak{B_c}+1)e^{2r^2W}+4\big)
+3Z(4\varpi\mathfrak{B_c}+1)e^{2r^2W}\\\nonumber
&+32\varpi\mathfrak{B_c}r^6W^3Z^2e^{4r^2W}+2r^4W^2Ze^{2r^2W}\big(-4(3\varpi\mathfrak{B_c}+1)+\varpi\mathfrak{B_c}Z^2e^{4r^2W}\\\nonumber
&+2Z(7\varpi\mathfrak{B_c}+1)e^{2r^2W}\big)-6\big)+\varpi
W^2\big(8(25\varpi\mathfrak{B_c}+16)+Z^2(24\varpi\mathfrak{B_c}+19)\\\nonumber
&\times
e^{4r^2W}+2r^2W\big(4(44\varpi\mathfrak{B_c}+31)+2Z^3(25\varpi\mathfrak{B_c}+8)e^{6r^2W}-Z^2(240\varpi\mathfrak{B_c}\\\nonumber
&+83)e^{4r^2W}-4Z(25\varpi\mathfrak{B_c}+23)e^{2r^2W}\big)-50Z(4\varpi\mathfrak{B_c}+3)e^{2r^2W}+384\varpi\mathfrak{B_c}\\\nonumber
&\times
r^{10}W^5Z^3e^{6r^2W}-32r^8W^4Z^2e^{4r^2W}\big(10\varpi\mathfrak{B_c}+\varpi\mathfrak{B_c}Z^2e^{4r^2W}-2Z(4\varpi\mathfrak{B_c}\\\nonumber
&+3)e^{2r^2W}+4\big)-8r^6W^3Ze^{2r^2W}\big(-16\varpi\mathfrak{B_c}+3\varpi\mathfrak{B_c}Z^3e^{6r^2W}-Z^2(33\varpi\mathfrak{B_c}\\\nonumber
&+10)e^{4r^2W}+220\varpi\mathfrak{B_c}Ze^{2r^2W}+12\big)+4r^4W^2\big(72\varpi\mathfrak{B_c}+Z^4(2\varpi\mathfrak{B_c}+1)\\\nonumber
&\times
e^{8r^2W}+Z^3(43\varpi\mathfrak{B_c}+9)e^{6r^2W}-2Z^2(166\varpi\mathfrak{B_c}+41)e^{4r^2W}-2Z(38\varpi\mathfrak{B_c}\\\label{g30}
&-13)e^{2r^2W}+12\big)\big)\bigg],
\end{align}
and the solution for $\mathbb{L}_m=P_r$ yields
\begin{align}\nonumber
\Delta&=\bigg[2\big(4\pi\big(r^2WZe^{2r^2W}+1\big)^2+\varpi
W\big(-Ze^{2r^2W}+4r^2W+2r^4W^2Ze^{2r^2W}\\\nonumber
&+3\big)\big)\big(32\pi\big(r^2WZe^{2r^2W}+1\big)^3-\varpi
W\big(2r^2W\big(10Z^2e^{4r^2W}-23Ze^{2r^2W}-6\big)\\\nonumber
&+17Ze^{2r^2W}+32r^6W^3Z^2e^{4r^2W}+4r^4W^2Ze^{2r^2W}\big(Z^2e^{4r^2W}+6\big)-50\big)\big)\bigg]^{-1}\\\nonumber
&\times\bigg[256\pi^2\mathfrak{B_c}\big(r^2WZe^{2r^2W}+1\big)^5-16\pi
W\big(r^2WZe^{2r^2W}+1\big)^2\big(-6(6\varpi\mathfrak{B_c}\\\nonumber
&+1)+r^2W\big(-8(3\varpi\mathfrak{B_c}+1)+Z^2(20\varpi\mathfrak{B_c}+3)e^{4r^2W}-2Z(32\varpi\mathfrak{B_c}+1)\\\nonumber
&\times
e^{2r^2W}\big)+Z(8\varpi\mathfrak{B_c}+3)e^{2r^2W}+24\varpi\mathfrak{B_c}r^6W^3Z^2e^{4r^2W}+2r^4W^2Ze^{2r^2W}\\\nonumber
&\times\big(-4(5\varpi\mathfrak{B_c}+1)+\varpi\mathfrak{B_c}Z^2e^{4r^2W}+2Z(3\varpi\mathfrak{B_c}+1)e^{2r^2W}\big)\big)+\varpi
W^2\big((33\\\nonumber
&\times\varpi\mathfrak{B_c}+16)8+Z^2(12\varpi\mathfrak{B_c}+19)e^{4r^2W}+2r^2W\big(4(56\varpi\mathfrak{B_c}+29)+2Z^3\\\nonumber
&\times(19\varpi\mathfrak{B_c}+8)e^{6r^2W}-Z^2(216\varpi\mathfrak{B_c}+83)e^{4r^2W}+8Z(55\varpi\mathfrak{B_c}-12)e^{2r^2W}\big)\\\nonumber
&-30Z(2\varpi\mathfrak{B_c}+5)e^{2r^2W}+128\varpi\mathfrak{B_c}r^{10}W^5Z^3e^{6r^2W}+16r^8W^4Z^2e^{4r^2W}\big(\varpi\mathfrak{B_c}\\\nonumber
&\times
Z^2e^{4r^2W}-48\varpi\mathfrak{B_c}+10Ze^{2r^2W}-8\big)+8r^6W^3Ze^{2r^2W}\big(-4(4\varpi\mathfrak{B_c}+3)\\\nonumber
&+5\varpi\mathfrak{B_c}Z^3e^{6r^2W}+Z^2(4\varpi\mathfrak{B_c}+9)e^{4r^2W}-2Z(46\varpi\mathfrak{B_c}+5)e^{2r^2W}\big)+4r^4W^2\\\nonumber
&\times\big(24\varpi\mathfrak{B_c}+Z^4(6\varpi\mathfrak{B_c}+1)e^{8r^2W}-Z^3(2\varpi\mathfrak{B_c}-9)e^{6r^2W}
-2Z^2(44\varpi\mathfrak{B_c}\\\label{g31}
&+43)e^{4r^2W}+10Z(26\varpi\mathfrak{B_c}+1)e^{2r^2W}+12\big)\big)\bigg].
\end{align}
We utilize the experimental data (Table $\mathbf{1}$) and calculated
constants (Table $\mathbf{2}$) to check the role of pressure
anisotropy in the development of compact structures. The anisotropy
shows increasing (outward) or decreasing (inward) behavior depending
on whether the tangential pressure is greater or lesser than the
radial pressure, respectively. Figure $\mathbf{5}$ (upper two plots)
shows that anisotropy in the interior of Cen X-3, Her X-I and RXJ
1856-37 stars varies from negative to positive, while it exhibits
decreasing behavior near the center and then increases towards
boundary inside 4U 1820-30 and 1808.4-3658 stars. This factor shows
same behavior for both values of $\varpi$.

According to Hossain \emph{et al.} \cite{41m}, negative anisotropy
allows the construction of massive stellar structure. It is
prominent from the graphical interpretation that anisotropy will
become positive after overcoming the negative value. The positive
anisotropy helps to construct the more compact object, according to
Gokhroo and Mehra \cite{41n}. Also, we can see that the anisotropy
increases and attains its maximum value at the boundary of each star
candidate which is an inherent property of an ultra dense compact
stars \cite{41o}.
\begin{figure}\center
\epsfig{file=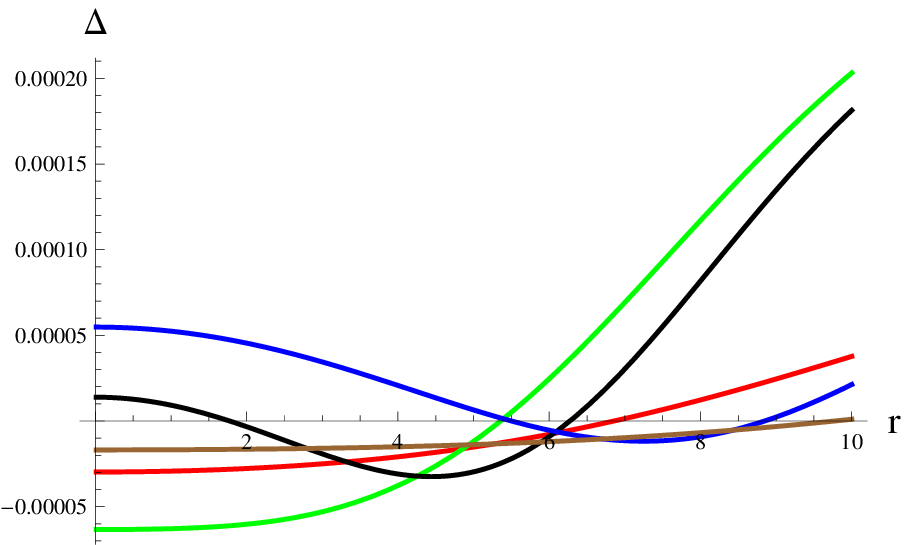,width=0.45\linewidth}\epsfig{file=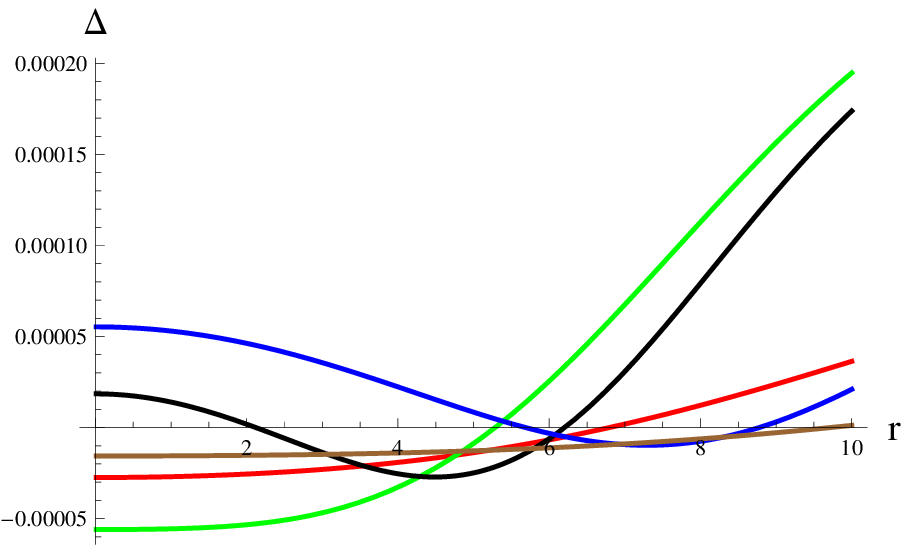,width=0.45\linewidth}
\epsfig{file=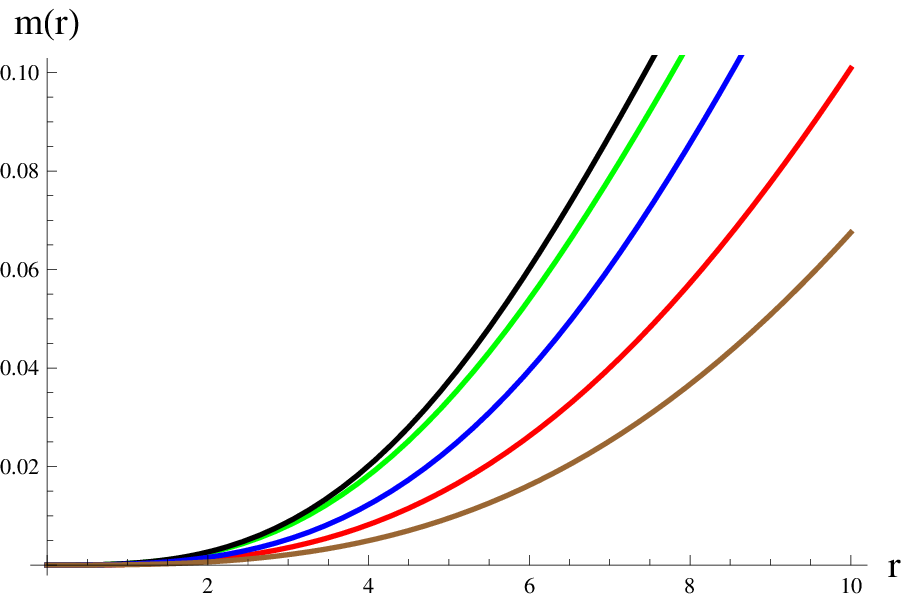,width=0.45\linewidth}\epsfig{file=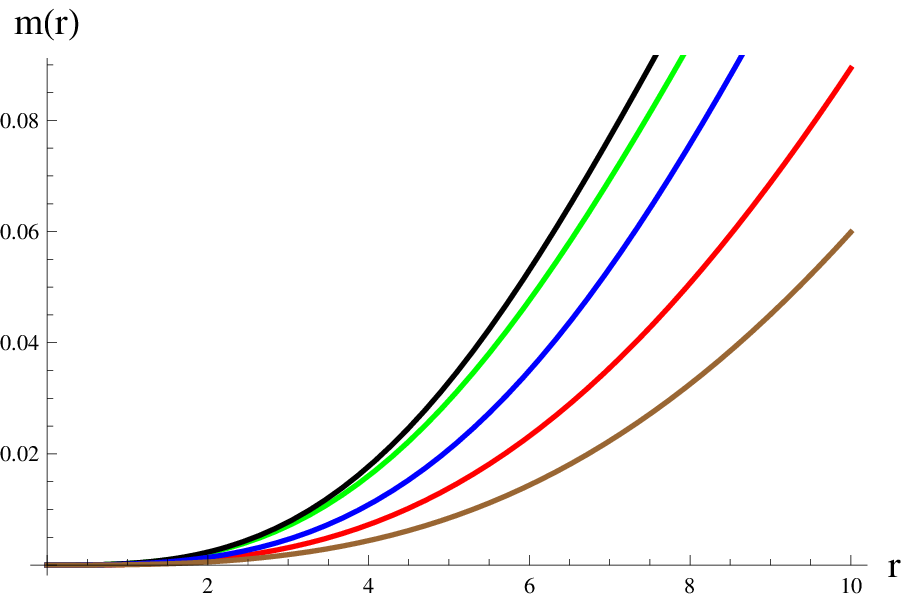,width=0.45\linewidth}
\caption{Variation of anisotropy and mass versus $r$ corresponding
to $\varpi=4$ and $\mathbb{L}_m=-\mu$ (left) as well as
$\mathbb{L}_m=P_r$ (right) for different compact star candidates}
\end{figure}

\subsection{Effective Mass, Compactness and Surface Redshift}

The effective mass of spherical geometry can be defined as
\begin{equation}\label{g32}
m(r)=\frac{1}{2}\int_{0}^{\mathcal{H}}r^2\mu dr,
\end{equation}
or it can be derived from Eq.\eqref{g12a} as
\begin{equation}\label{g33}
m(r)=\frac{r}{2}\left\{\frac{2\bar{M}r^2e^{\frac{\bar{M}\left(\mathcal{H}^2-r^2\right)}{\mathcal{H}^2(2\bar{M}-\mathcal{H})}}}
{2\bar{M}r^2e^{\frac{\bar{M}\left(\mathcal{H}^2-r^2\right)}{\mathcal{H}^2(2\bar{M}-\mathcal{H})}}+\mathcal{H}^2(\mathcal{H}-2\bar{M})}\right\},
\end{equation}
which shows that there is no mass at the center of each strange
star. Figure $\mathbf{5}$ (second row) contains the plots of mass
for both solutions which show that compact objects become more
massive for the first solution corresponding to $\mathbb{L}_m=-\mu$.
The evolution of massive bodies can be studied by investigating
various physical quantities, i.e., the compactness factor which is
defined as the mass to radius ratio of a compact star. Its
mathematical expression is given as
\begin{equation}\label{g34}
\sigma(r)=\frac{m(r)}{r}=\frac{1}{2}\left\{\frac{2\bar{M}r^2e^{\frac{\bar{M}\left(\mathcal{H}^2-r^2\right)}{\mathcal{H}^2(2\bar{M}-\mathcal{H})}}}
{2\bar{M}r^2e^{\frac{\bar{M}\left(\mathcal{H}^2-r^2\right)}{\mathcal{H}^2(2\bar{M}-\mathcal{H})}}+\mathcal{H}^2(\mathcal{H}-2\bar{M})}\right\},
\end{equation}
whose maximum value for a feasible solution corresponding to a
self-gravitating body has been found by Buchdahl \cite{42a} after
matching the interior and exterior regions of spacetimes at
hypersurface ($r=\mathcal{H}$) as $\frac{4}{9}$. The measurement of
wavelength of electromagnetic radiations emitting from a massive
object with enough gravitational attraction is known as redshift,
given as
\begin{equation}\label{g35}
D(r)=\frac{1}{\sqrt{1-2\sigma(r)}}-1,
\end{equation}
which further becomes
\begin{equation}\label{g36}
D(r)=-1+\sqrt{\frac{2\bar{M}r^2e^{\frac{\bar{M}\left(\mathcal{H}^2-r^2\right)}{\mathcal{H}^2(2\bar{M}-\mathcal{H})}}+\mathcal{H}^2(\mathcal{H}
-2\bar{M})}{\mathcal{H}^2(\mathcal{H}-2\bar{M})}}.
\end{equation}
Buchdahl proposed its upper limit inside a feasible compact star
having perfect fluid as $2$, while it was observed to be $5.211$
\cite{42b} for anisotropic configured structures. Figure
$\mathbf{6}$ shows the plots of these factors with respect to each
star for both solutions and we find them consistent with their
respective limits for $\varpi=4$ (Tables $\mathbf{3}$ and
$\mathbf{4}$). It is seen that increment in the value of bag
constant increases the above factors. The behavior of mass,
compactness and redshift meets their respective acceptability
criteria for $\varpi=-4$ as well.
\begin{figure}\center
\epsfig{file=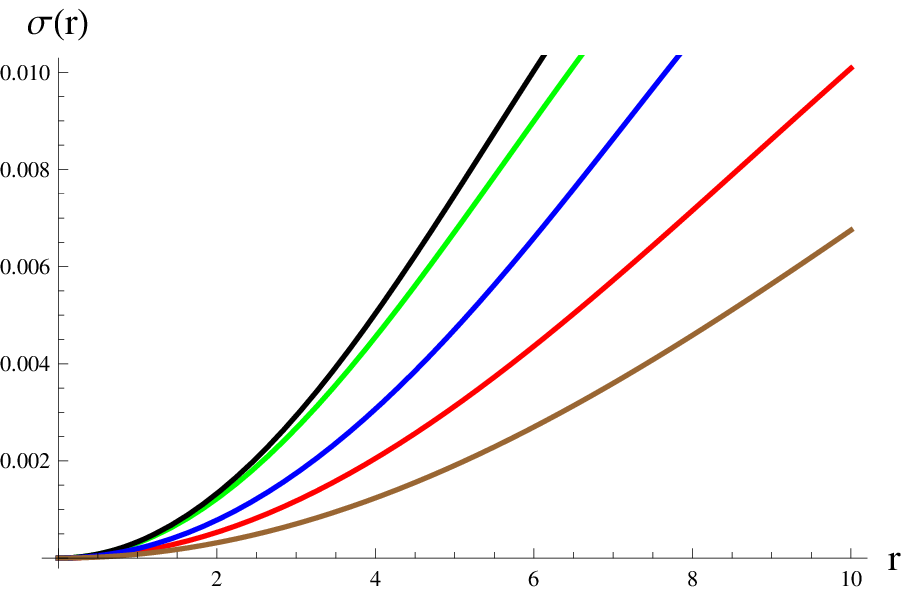,width=0.45\linewidth}\epsfig{file=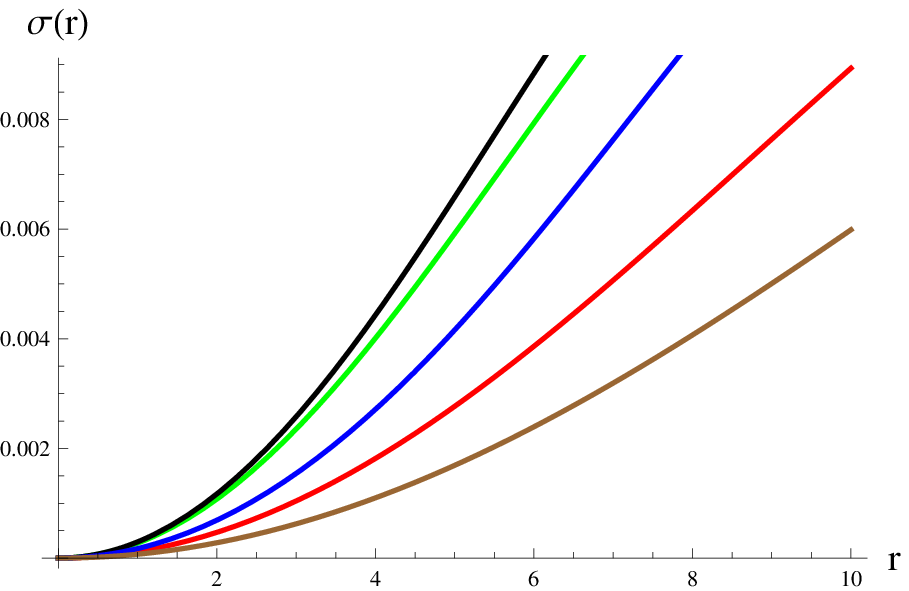,width=0.45\linewidth}
\epsfig{file=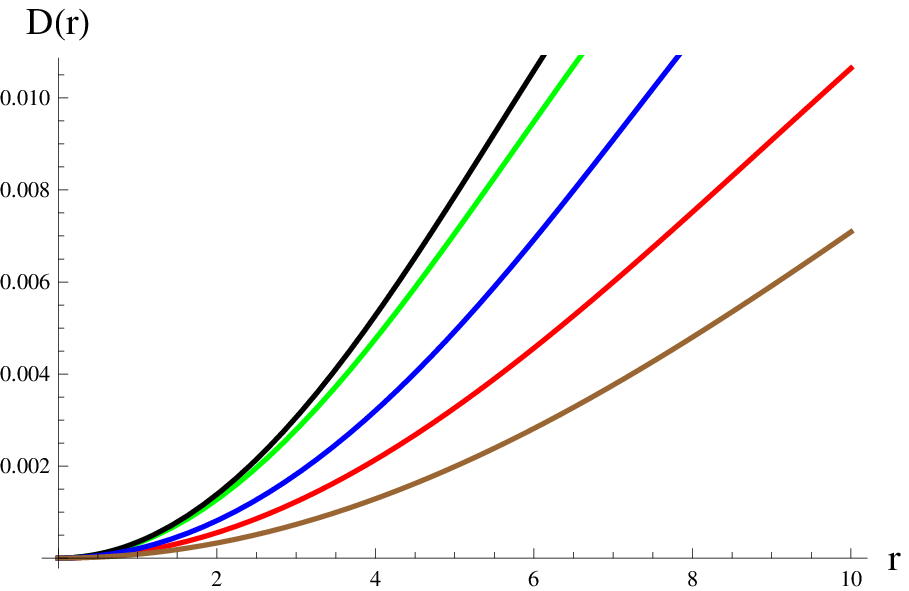,width=0.45\linewidth}\epsfig{file=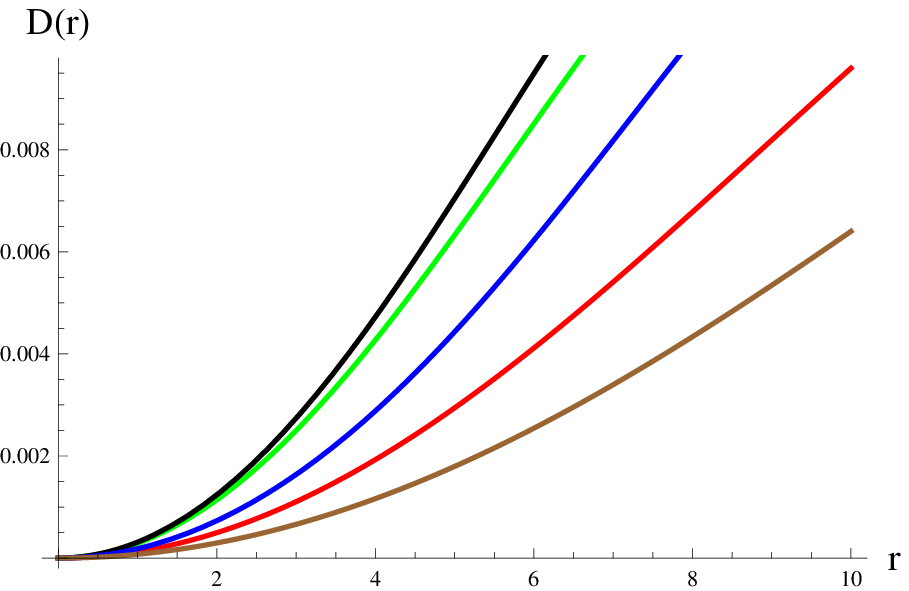,width=0.45\linewidth}
\caption{Variation of compactness and redshift factors versus $r$
corresponding to $\varpi=4$ and $\mathbb{L}_m=-\mu$ (left) as well
as $\mathbb{L}_m=P_r$ (right) for different compact star candidates}
\end{figure}

\subsection{Energy Conditions}

Various constrains involving the state variables (and charge in the
presence of electromagnetic field) have widely been used in
astrophysics to ensure whether the matter in a particular geometry
is normal or exotic. The fulfilment of such bounds also guarantee
viability of the resulting solution. A realistic configuration is
the one which satisfies all the following constraints
\begin{itemize}
\item Null: $\mu+P_r \geq 0$, \quad $\mu+P_\bot \geq 0$,
\item Weak: $\mu \geq 0$, \quad $\mu+P_r \geq 0$, \quad $\mu+P_\bot \geq 0$,
\item Strong: $\mu+P_r+2P_\bot \geq 0$,
\item Dominant: $\mu-P_r \geq 0$, \quad $\mu-P_\bot \geq 0$.
\end{itemize}
The graphical analysis of the above conditions for both resulting
solutions is presented in Figures $\mathbf{7}$ and $\mathbf{8}$,
from which it is observed that our developed solutions as well as
$f(\mathcal{R},\mathcal{T},\mathcal{Q})$ model \eqref{g61} are
physically viable for $\varpi=4$. We also check these conditions
with respect to $\varpi=-4$ and obtain viable results, but their
graphs have not been added. Thus it can be doubtlessly said that
there must exist normal matter in the interior of all quark
candidates.
\begin{figure}\center
\epsfig{file=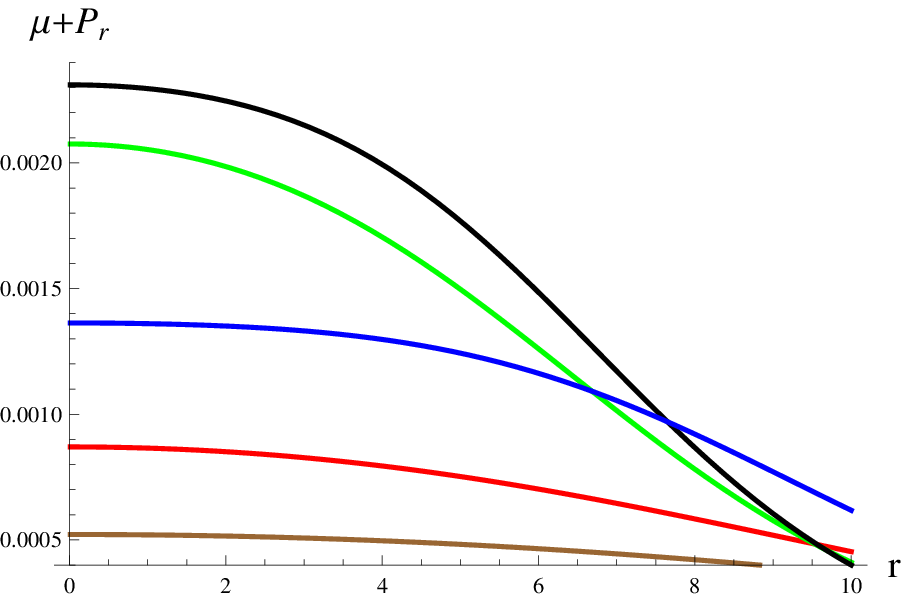,width=0.45\linewidth}\epsfig{file=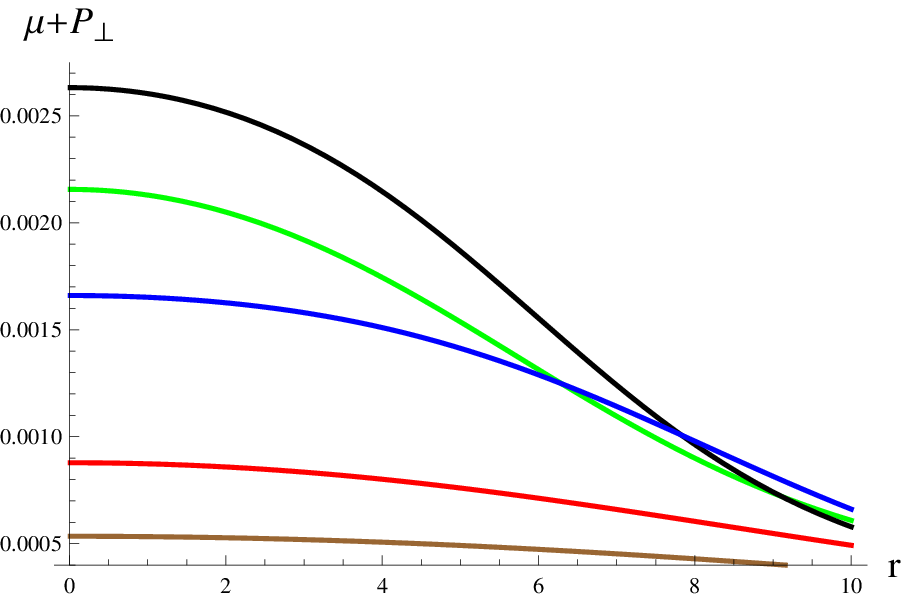,width=0.45\linewidth}
\epsfig{file=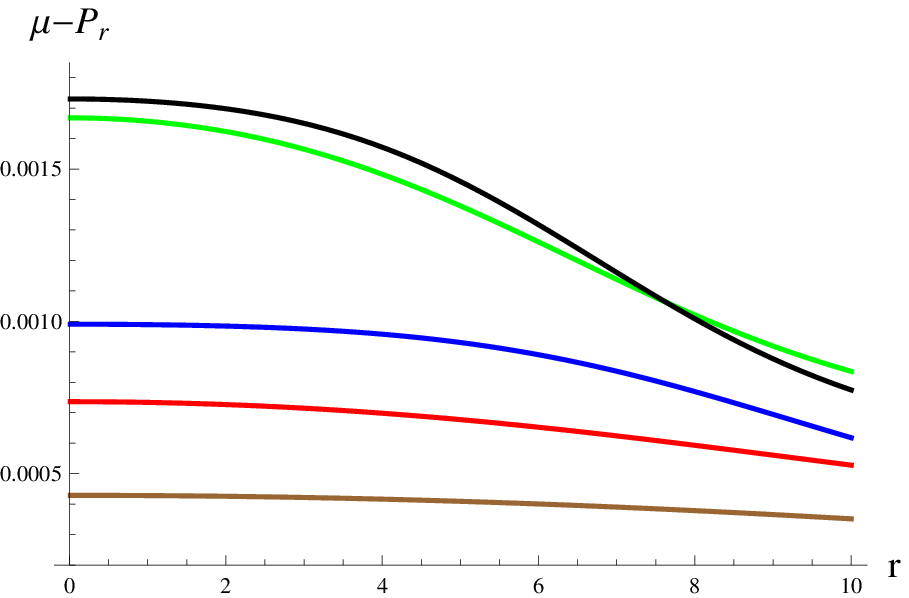,width=0.45\linewidth}\epsfig{file=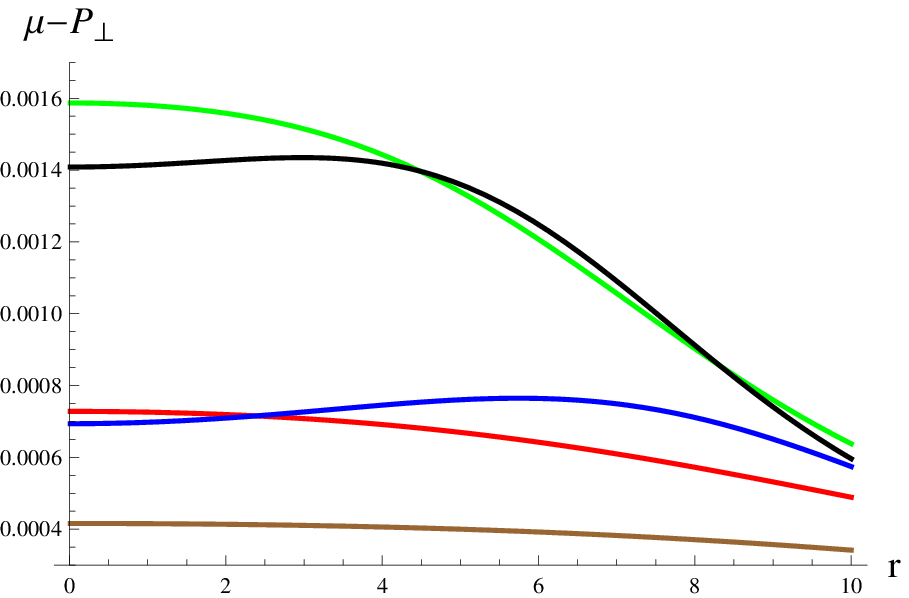,width=0.45\linewidth}
\epsfig{file=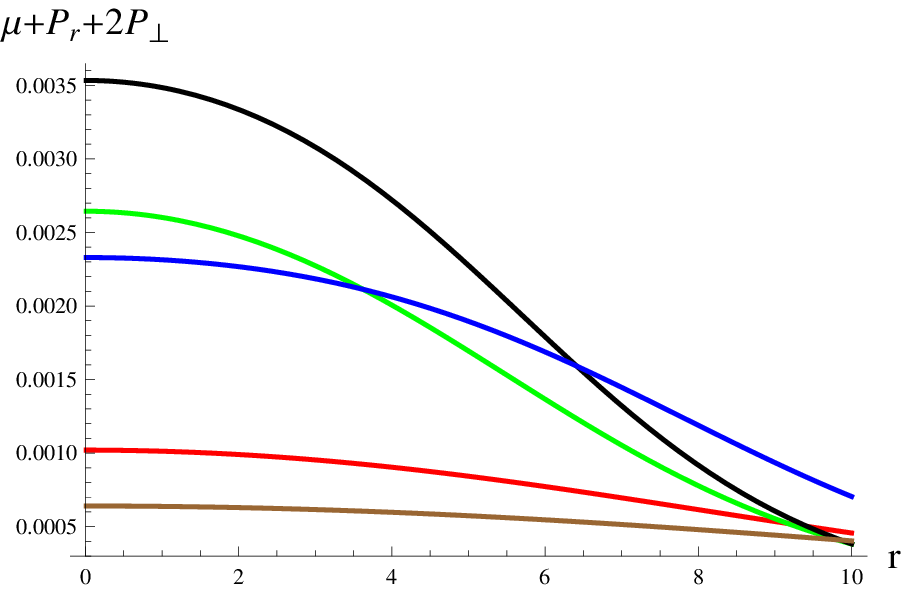,width=0.45\linewidth} \caption{Plots of
energy conditions versus $r$ corresponding to $\varpi=4$ and
$\mathbb{L}_m=-\mu$ for different compact star candidates}
\end{figure}
\begin{figure}\center
\epsfig{file=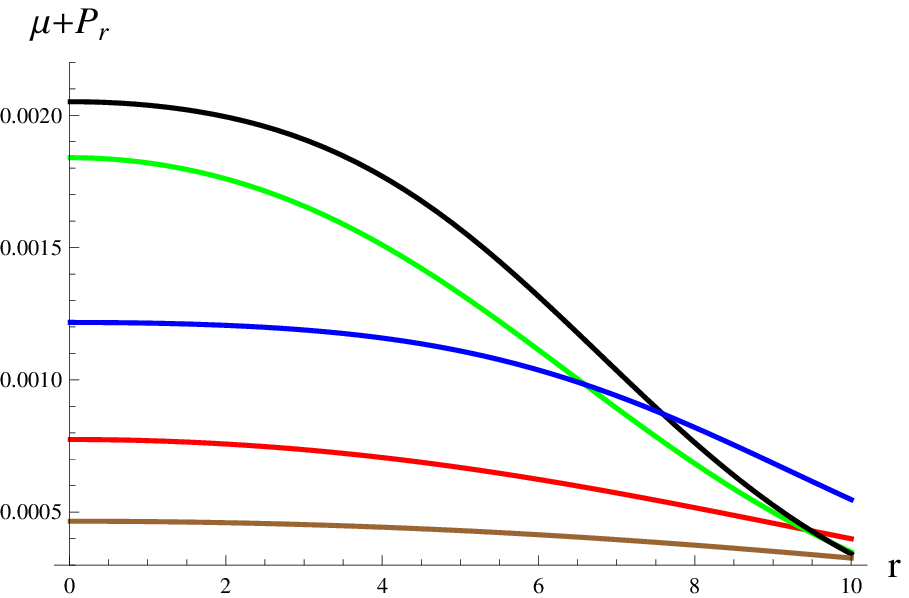,width=0.45\linewidth}\epsfig{file=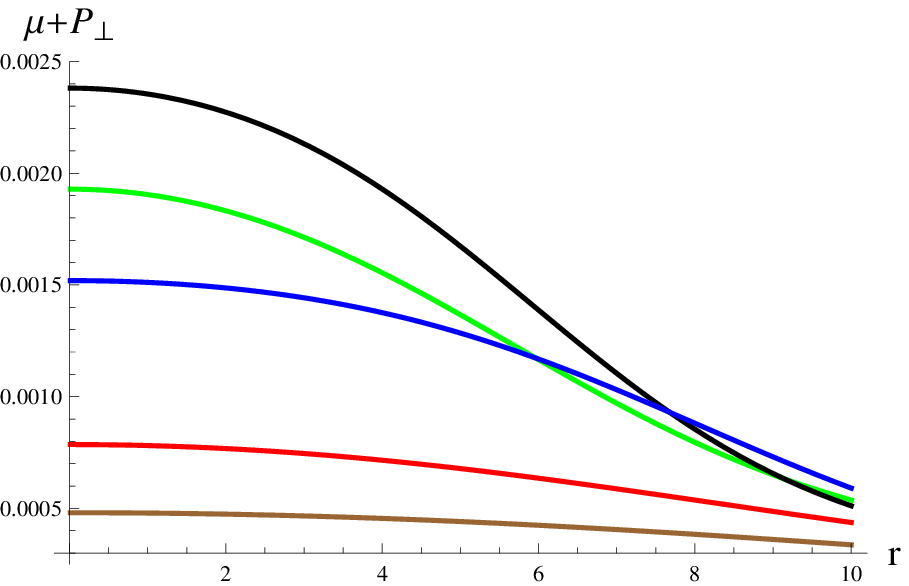,width=0.45\linewidth}
\epsfig{file=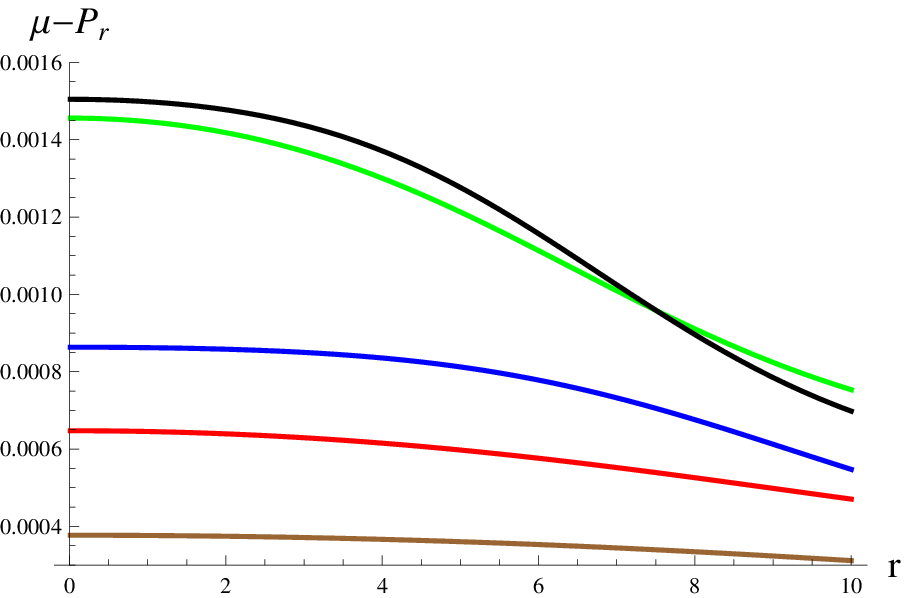,width=0.45\linewidth}\epsfig{file=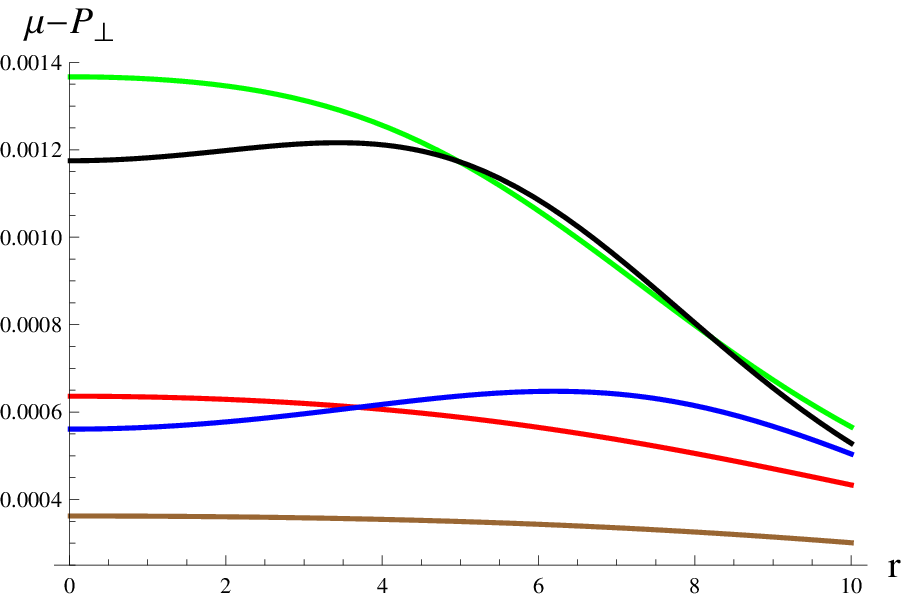,width=0.45\linewidth}
\epsfig{file=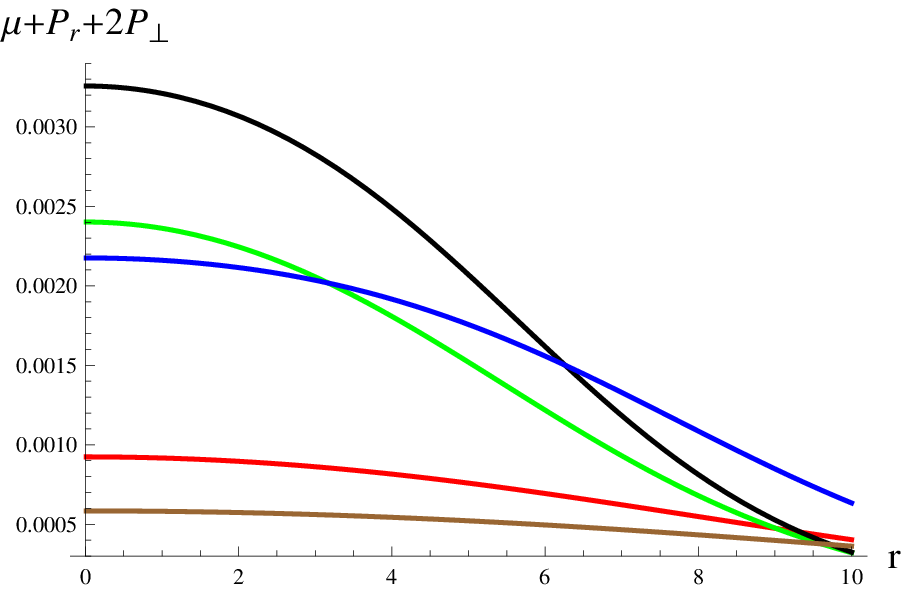,width=0.45\linewidth} \caption{Plots of
energy conditions versus $r$ corresponding to $\varpi=4$ and
$\mathbb{L}_m=P_r$ for different compact star candidates}
\end{figure}

\subsection{Stability Analysis}

The stability of physical models and astronomical objects gained
much attention in explaining different phases of our cosmos. One can
get better understanding about the structural evolution of compact
bodies which meet the stability criteria. In this regard, various
approaches have been discussed in literature such as causality
condition and adiabatic index etc. We utilize these techniques to
analyze the stability of quark candidates for the model \eqref{g61}.
According to the causality condition \cite{42d}, the speed of light
must be greater than the speed of sound within a stable system,
i.e., $0 < v_{sr}^{2} < 1$ and $0 < v_{s\bot}^{2} < 1$. Here,
$v_{sr}$ and $v_{s\bot}$ are sound speeds in radial and tangential
directions, respectively and expressed as
\begin{equation}
v_{sr}^{2}=\frac{dP_{r}}{d\mu}, \quad
v_{s\bot}^{2}=\frac{dP_{\bot}}{d\mu}.
\end{equation}
The stability can also be checked with the help of Herrera's
cracking concept \cite{29i} which states that stable structure must
satisfy the inequality $0 < \mid v_{s\bot}^{2}-v_{sr}^{2} \mid < 1$
in its interior. The gravitational cracking happens when the radial
force is directed inwards in the inner part of the sphere for all
values of the radial coordinate $r$ between the center, and some
value beyond which the force reverses its direction.

Another powerful approach which helps to analyze the stability of
self-gravitating structure is the adiabatic index
$\big(\Gamma\big)$. Many researchers \cite{42e,42g} utilized this
tool to study celestial bodies and concluded that its value should
not be less than $\frac{4}{3}$ everywhere in stable models. The
mathematical representation of $\Gamma$ is
\begin{equation}\label{g62}
\Gamma=\frac{\mu+P_{r}}{P_{r}}
\left(\frac{dP_{r}}{d\mu}\right)=\frac{\mu+P_{r}}{P_{r}}
\left(v_{sr}^{2}\right).
\end{equation}
Figure $\mathbf{9}$ depicts the graphs of $\mid
v_{s\bot}^{2}-v_{sr}^{2} \mid$ and $\Gamma$ for each quark star
corresponding to both solutions for $\varpi=4$. The upper left plot
shows that all stars are stable everywhere except 4U 1820-30 (which
is unstable near its core and stable towards boundary) with respect
to the solution for $\mathbb{L}_m=-\mu$, while the solution
corresponding to $\mathbb{L}_m=P_r$ provides that two stars, namely
4U 1820-30 and SAX J 1808.4-3658 are unstable near their center
(upper right plot).

As $\varpi=-4$ is concerned, the stability of the solution
corresponding to $\mathbb{L}_m=-\mu$ produces same results as we
have obtained for $\varpi=4$. On the contrary, the interior of the
compact star SAX J 1808.4-3658 is stable throughout corresponding to
the solution with respect to $\mathbb{L}_m=P_r$, as shown in Figure
$\mathbf{10}$. The adiabatic index is found to be within its
acceptable range for both the cases (lower plots of Figures
$\mathbf{9}$ and $\mathbf{10}$).
\begin{figure}\center
\epsfig{file=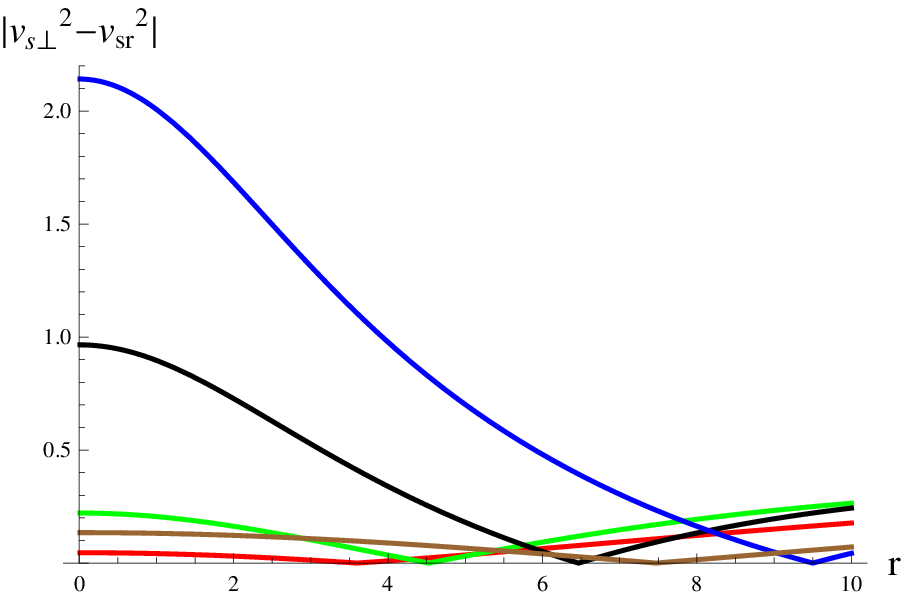,width=0.45\linewidth}\epsfig{file=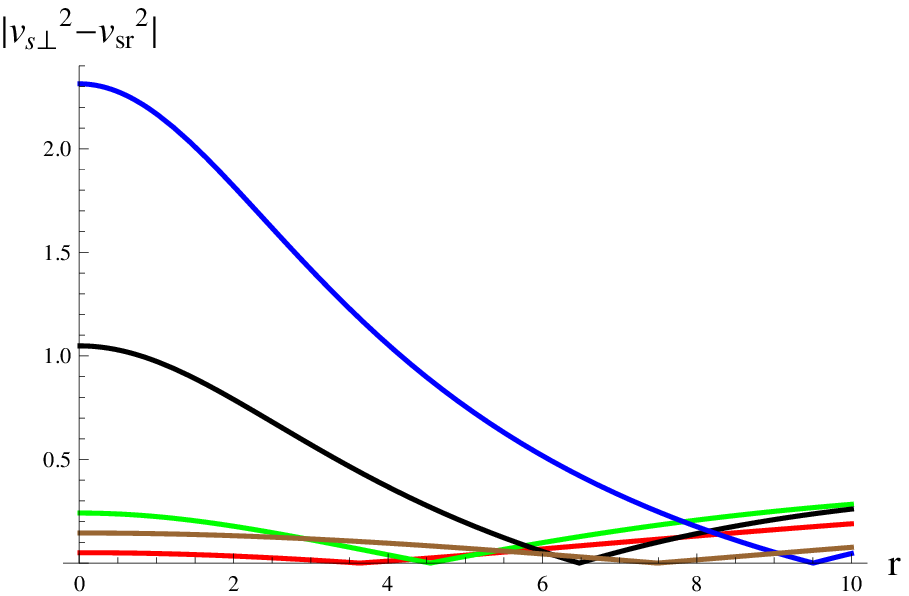,width=0.45\linewidth}
\epsfig{file=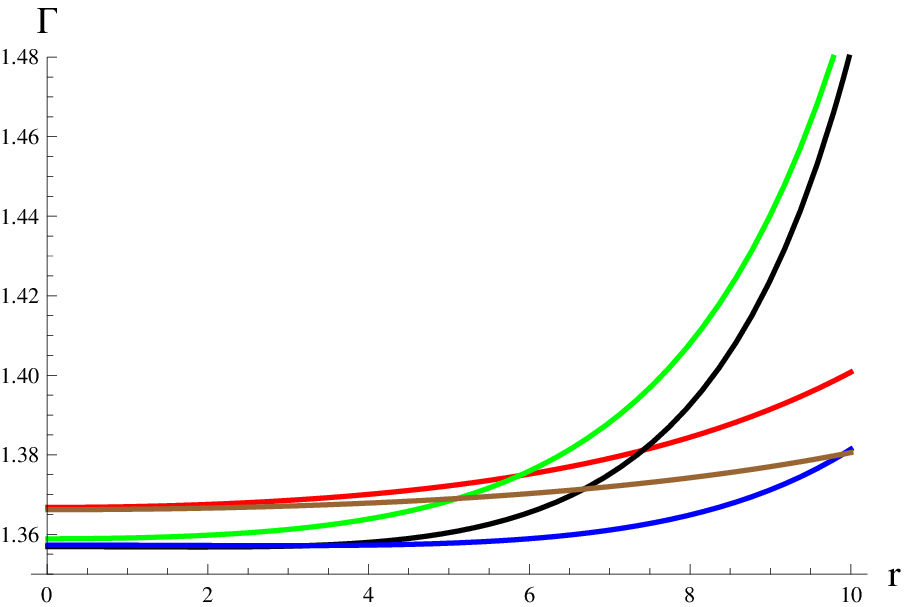,width=0.45\linewidth}\epsfig{file=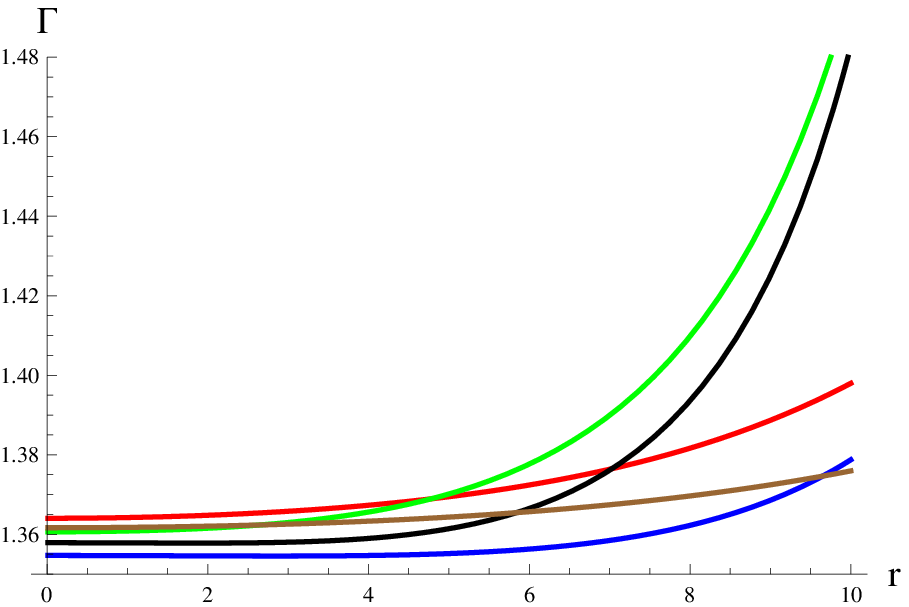,width=0.45\linewidth}
\caption{Plots of $\mid v_{s\bot}^{2}-v_{sr}^{2} \mid$ and adiabatic
index versus $r$ corresponding to $\varpi=4$ and $\mathbb{L}_m=-\mu$
(left) as well as $\mathbb{L}_m=P_r$ (right) for different compact
star candidates}
\end{figure}
\begin{figure}\center
\epsfig{file=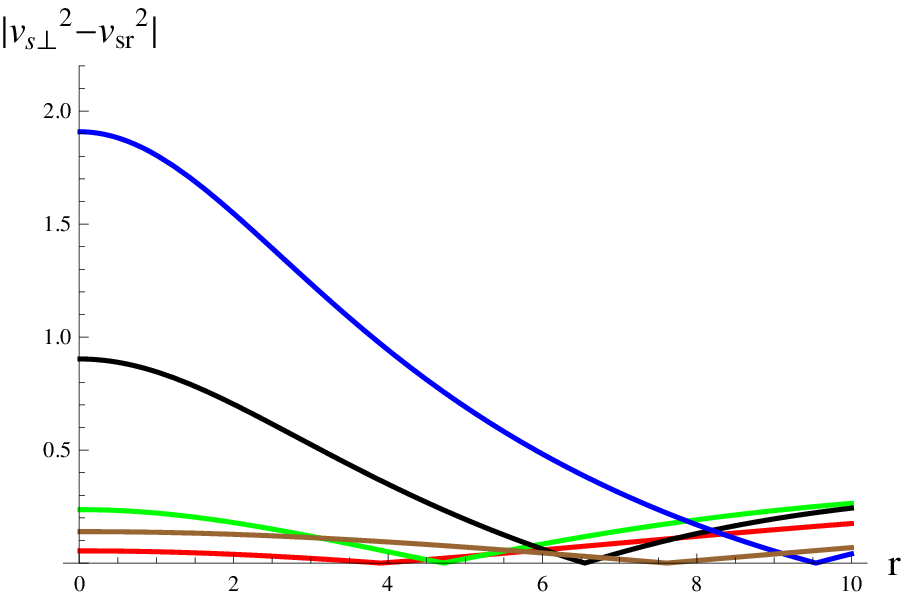,width=0.45\linewidth}\epsfig{file=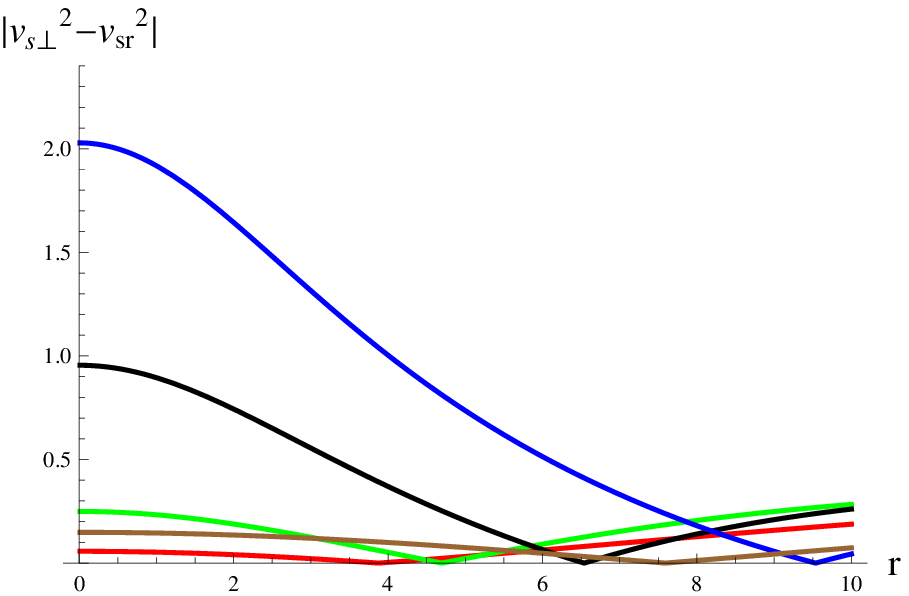,width=0.45\linewidth}
\epsfig{file=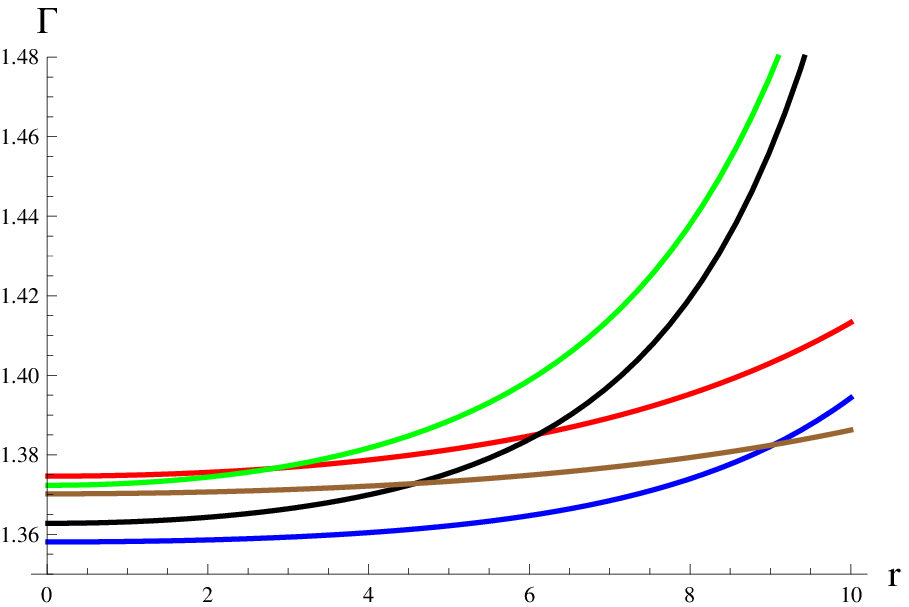,width=0.45\linewidth}\epsfig{file=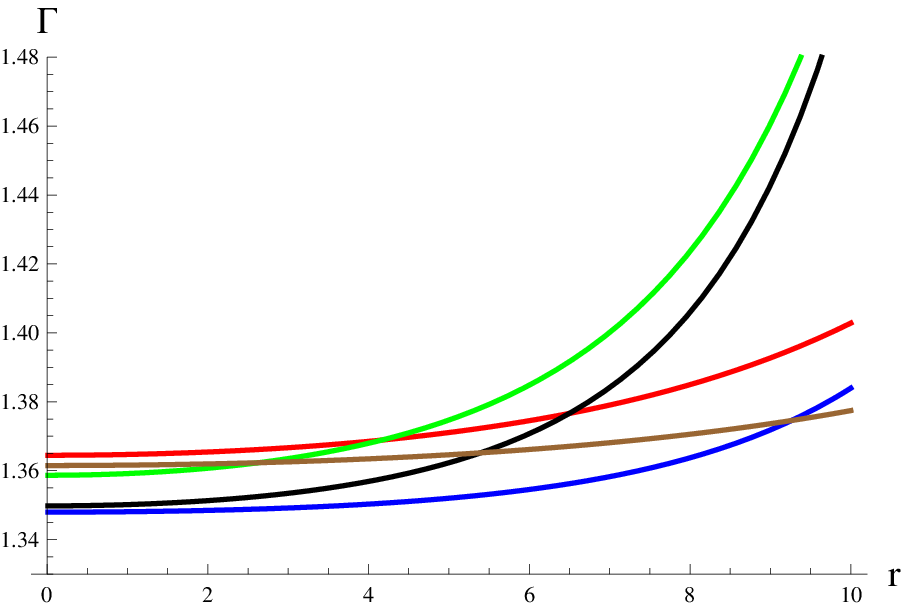,width=0.45\linewidth}
\caption{Plots of $\mid v_{s\bot}^{2}-v_{sr}^{2} \mid$ and adiabatic
index versus $r$ corresponding to $\varpi=-4$ and
$\mathbb{L}_m=-\mu$ (left) as well as $\mathbb{L}_m=P_r$ (right) for
different compact star candidates}
\end{figure}

\section{Final Remarks}

This paper explores the existence of anisotropic stars in the
framework of
$f(\mathcal{R},\mathcal{T},\mathcal{R}_{\omega\alpha}\mathcal{T}^{\omega\alpha})$
gravity. We have taken a linear model $\mathcal{R}+\varpi
\mathcal{R}_{\omega\alpha}\mathcal{T}^{\omega\alpha}$ which still
upholds the matter-geometry coupling effects in the interior of
compact stars, where the coupling constant $\varpi$ has been chosen
as $\pm4$. The modified field equations and $\mathbb{TOV}$ equation
have been formulated for the above model with respect to two
different choices of matter Lagrangian. We have adopted an
acceptable temporal metric function \cite{37,37a} and utilized
embedding class-one condition to calculate radial metric potential
\eqref{g15} and found a solution to the field equations. We have
also assumed $\mathbb{MIT}$ bag model $\mathbb{E}$o$\mathbb{S}$
which interconnects the energy density and radial pressure of the
inner configuration. Both metric coefficients involve four unknowns
($W,X,Y,Z$) whose values have been calculated at the boundary in
terms of mass and radius of celestial body. The observational data
of five different strange stars, i.e., 4U 1820-30,~Cen X-3,~RXJ
1856-37,~SAX J 1808.4-3658 and Her X-I (Table $\mathbf{1}$) have
been employed to calculate unknown quantities (Table $\mathbf{2}$)
and bag constant with respect to each candidate. Tables $\mathbf{3}$
and $\mathbf{4}$ contain the values of energy density (at center and
surface), radial pressure (at center), compactness as well as
redshift (at surface) and bag constant for both solutions. Figure
$\mathbf{5}$ shows that mass of each quark body decreases by
decreasing the bag constant. The graphical interpretation of matter
variables has also been analyzed. It is found that the state
variables corresponding to both solutions show physically acceptable
behavior as they are maximum at the center and minimum at the
boundary of respective star candidate.

The interior of all stars become more dense for the solution
corresponding to $\mathbb{L}_m=-\mu$, whereas $\mathbb{L}_m=P_r$
provides less dense structures. We have found acceptable behavior of
redshift and compactness (Figure $\mathbf{6}$). The positive
behavior of energy bounds confirmed the viability of both developed
solutions as well as presence of normal matter in the interior of
stellar bodies (Figures $\mathbf{7}$ and $\mathbf{8}$). Finally, we
have utilized two approaches to check the stability of resulting
solutions for $\varpi=\pm4$. The first one is the Herrera's cracking
approach which guarantees the fulfilment of the inequality $0<\mid
v_{s\bot}^{2}-v_{sr}^{2} \mid<1$ within stable system. All candidate
stars are observed to be stable except 4U 1820-30 (which is unstable
near its core) for the solution corresponding to $\mathbb{L}_m=-\mu$
(upper left plot of Figures $\mathbf{9}$ and $\mathbf{10}$), while
for $\mathbb{L}_m=P_r$ and $\varpi=4$, stars 4U 1820-30 and SAX J
1808.4-3658 become unstable near their center and show stable
behavior towards the boundary (upper right plot of Figure
$\mathbf{9}$). However, the compact candidate SAX J 1808.4-3658 is
observed to be stable throughout for the solution with respect to
$\mathbb{L}_m=P_r$ and $\varpi=-4$ (upper right plot of Figure
$\mathbf{10}$). The other three stars, namely Cen X-3,~Her X-I and
RXJ 1856-37 are found to be stable with respect to each solution.
Figures $\mathbf{9}$ and $\mathbf{10}$ (lower plots) show the
behavior of adiabatic index which provides acceptable values
everywhere.

In the framework of $\mathbb{GR}$, the central density, surface
density as well as central radial pressure corresponding to two
different stars, namely Her X-1 and RXJ 1856-37 have been calculated
\cite{37}. By comparing our results, we observe that these physical
quantities have less values in this modified gravity. The values of
physical variables inside the quark star SAX J 1808.4-3658 have been
determined in $f(\mathcal{R},\mathcal{T})$ theory \cite{34aa}, from
which we have found that the internal configuration of this star
becomes more dense in $f(\mathcal{R},\mathcal{T},\mathcal{Q})$. The
interior of the star candidate 4U 1820-30 in this theory is found to
be less dense than that in $f(\mathcal{G})$ gravity \cite{38g}. For
$\varpi=-4$, we obtain more suitable results in
$f(\mathcal{R},\mathcal{T},\mathcal{R}_{\omega\alpha}\mathcal{T}^{\omega\alpha})$
theory as compared to [25] as well as the solution corresponding to
$\varpi=4$. Finally, one can reduce all these results to
$\mathbb{GR}$ by taking $\varpi=0$ in the modified model
\eqref{g61}.

\vspace{0.25cm}

\section*{Appendix A}

The state variables \eqref{g14b}-\eqref{g14d} in terms of unknowns
$(W,X,Y,Z)$ take the form
\begin{align}\nonumber
\mu&=\bigg[32\pi\big(r^2WZe^{2r^2W}+1\big)^3-\varpi
W\big\{2r^2W\big(14Z^2e^{4r^2W}+Ze^{2r^2W}-6\big)+25Z\\\nonumber
&\times
e^{2r^2W}+32r^6W^3Z^2e^{4r^2W}+4r^4W^2Ze^{2r^2W}\big(Z^2e^{4r^2W}+8Ze^{2r^2W}+6\big)-34\big\}\bigg]^{-1}\\\nonumber
&\times\bigg[32\pi\mathfrak{B_c}+2r^2W^2\big(24\varpi\mathfrak{B_c}+Z^2e^{4r^2W}\big(-20\varpi\mathfrak{B_c}
+48\pi\mathfrak{B_c}r^2+3\big)-2Z\\\nonumber
&\times(14\varpi\mathfrak{B_c}-9)e^{2r^2W}\big)+2W\big(20\varpi\mathfrak{B_c}+Ze^{2r^2W}\big(-20\varpi\mathfrak{B_c}
+48\pi\mathfrak{B_c}r^2+3\big)\\\nonumber
&+6\big)-32\varpi\mathfrak{B_c}r^6W^4Z^2e^{4r^2W}+4r^4W^3Z^2e^{4r^2W}\big(-20\varpi\mathfrak{B_c}+\mathfrak{B_c}Z\big(8\pi
r^2-\varpi\big)\\\nonumber &\times
e^{2r^2W}+6\big)\bigg],\\\nonumber
P_r&=\bigg[32\pi\big(r^2WZe^{2r^2W}+1\big)^3-\varpi
W\big\{2r^2W\big(14Z^2e^{4r^2W}+Ze^{2r^2W}-6\big)+25Z\\\nonumber
&\times
e^{2r^2W}+32r^6W^3Z^2e^{4r^2W}+4r^4W^2Ze^{2r^2W}\big(Z^2e^{4r^2W}+8Ze^{2r^2W}+6\big)-34\big\}\bigg]^{-1}\\\nonumber
&\times\bigg[-2\big(r^2WZe^{2r^2W}+1\big)\big(16\pi\mathfrak{B_c}+2r^2W^2Ze^{2r^2W}\big(-4\varpi\mathfrak{B_c}+\big(8\pi
r^2-\varpi\big)\\\nonumber
&\mathfrak{B_c}Ze^{2r^2W}-2\big)+W\big(16\varpi\mathfrak{B_c}+Ze^{2r^2W}\big(-10\varpi\mathfrak{B_c}
+32\pi\mathfrak{B_c}r^2-1\big)-2\big)-16\\\nonumber
&\times\varpi\mathfrak{B_c}r^4W^3Ze^{2r^2W}\big)\bigg],\\\nonumber
P_\bot&=\bigg[2\big\{\varpi
W\big(-Ze^{2r^2W}+2r^2W+3\big)+4\pi\big(r^2WZe^{2r^2W}+1\big)^2\big\}\big\{32\pi\big(r^2WZe^{2r^2W}\\\nonumber
&+1\big)^3-\varpi
W\big(2r^2W\big(14Z^2e^{4r^2W}+Ze^{2r^2W}-6\big)+25Ze^{2r^2W}+32r^6W^3Z^2e^{4r^2W}\\\nonumber
&+4r^4W^2Ze^{2r^2W}\big(Z^2e^{4r^2W}+8Ze^{2r^2W}+6\big)-34\big)\big\}\bigg]^{-1}\bigg[W\big\{\varpi
W\big(8(\varpi\mathfrak{B_c}+19)\\\nonumber
&-Z^2(16\varpi\mathfrak{B_c} -15)e^{4 r^2 W}+2 r^2 W \big(4 (28
\varpi \mathfrak{B_c} +33)+2 Z^3 (13 \varpi  \mathfrak{B_c} +7) e^{6
r^2 W}\\\nonumber &-Z^2 (152 \varpi \mathfrak{B_c} +89) e^{4 r^2
W}-4 Z (27 \varpi \mathfrak{B_c} +13) e^{2 r^2 W}\big)-2 Z (8 \varpi
\mathfrak{B_c} +73) e^{2 r^2 W}\\\nonumber &+384 \varpi
\mathfrak{B_c} r^{10} W^5 Z^3 e^{6 r^2 W}-32 r^8 W^4 Z^2 e^{4 r^2 W}
\big(6 \varpi \mathfrak{B_c} +\varpi \mathfrak{B_c} Z^2 e^{4 r^2
W}-2 Z (4 \varpi \mathfrak{B_c}\\\nonumber &+3) e^{2 r^2 W}+4\big)-8
r^6 W^3 Z e^{2 r^2 W} \big(-32 \varpi \mathfrak{B_c} +3 \varpi
\mathfrak{B_c} Z^3 e^{6 r^2 W}-Z^2 (27 \varpi \mathfrak{B_c}
+10)\\\nonumber &\times e^{4 r^2 W}+4 Z (47 \varpi \mathfrak{B_c}
-1) e^{2 r^2 W}+12\big)+4 r^4 W^2 \big(72 \varpi \mathfrak{B_c} +Z^4
e^{8 r^2 W}+Z^3e^{6 r^2 W}\\\nonumber &(41 \varpi \mathfrak{B_c}
+5)-4 Z^2 (75 \varpi \mathfrak{B_c} +17) e^{4 r^2 W}-2 Z (22 \varpi
\mathfrak{B_c} -19) e^{2 r^2 W}+12\big)\big)-32 \pi\\\nonumber
&\times\big(r^2 W Z e^{2 r^2 W}+1\big)^2 \big(r^2 W \big(-4 (\varpi
\mathfrak{B_c} +1)+Z^2 (4 \varpi \mathfrak{B_c} +1) e^{4 r^2 W}-2
Ze^{2 r^2 W}\\\nonumber &\times(5 \varpi \mathfrak{B_c}
+2)\big)+e^{2 r^2 W} (Z-\varpi \mathfrak{B_c}  Z)+8 \varpi
\mathfrak{B_c}  r^6 W^3 Z^2 e^{4 r^2 W}+2 r^4 W^2 Z e^{2 r^2
W}\\\nonumber &\times\big(\varpi \mathfrak{B_c} \big(5 Z e^{2 r^2
W}-8\big)-2\big)-4\big)\big\}\bigg],
\end{align}
and Eqs.\eqref{g14e}-\eqref{g14g} become
\begin{align}\nonumber
\mu&=\bigg[32\pi\big(r^2WZe^{2r^2W}+1\big)^3-\varpi
W\big\{2r^2W\big(10Z^2e^{4r^2W}-23Ze^{2r^2W}-6\big)+17Z\\\nonumber
&\times
e^{2r^2W}+32r^6W^3Z^2e^{4r^2W}+4r^4W^2Ze^{2r^2W}\big(Z^2e^{4r^2W}+6\big)-50\big\}\bigg]^{-1}\bigg[32\pi\mathfrak{B_c}\\\nonumber
&+2r^2W^2\big(24\varpi\mathfrak{B_c}
+Z^2e^{4r^2W}\big(-16\varpi\mathfrak{B_c}+48\pi\mathfrak{B_c}r^2+3\big)-2Z(2\varpi\mathfrak{B_c}-9)e^{2r^2W}\big)\\\nonumber
&+2W\big(28\varpi\mathfrak{B_c}+Ze^{2r^2W}\big(-16\varpi\mathfrak{B_c}+48\pi\mathfrak{B_c}r^2+3\big)+6\big)
-32\varpi\mathfrak{B_c}r^6W^4Z^2e^{4r^2W}\\\nonumber
&+4r^4W^3Z^2e^{4r^2W}\big(-12\varpi\mathfrak{B_c}+\mathfrak{B_c}Z\big(8\pi
r^2-\varpi\big)e^{2r^2W}+6\big)\bigg],\\\nonumber
P_r&=\bigg[32\pi\big(r^2WZe^{2r^2W}+1\big)^3-\varpi
W\big\{2r^2W\big(10Z^2e^{4r^2W}-23Ze^{2r^2W}-6\big)+17Z\\\nonumber
&\times
e^{2r^2W}+32r^6W^3Z^2e^{4r^2W}+4r^4W^2Ze^{2r^2W}\big(Z^2e^{4r^2W}+6\big)-50\big\}\bigg]^{-1}\bigg[-2\big(r^2\\\nonumber
&\times WZe^{2r^2W}+1\big)\big(16\pi\mathfrak{B_c}
+2r^2W^2Ze^{2r^2W}\big(4\varpi\mathfrak{B_c}+\mathfrak{B_c}Z\big(8\pi
r^2-\varpi\big)e^{2r^2W}-2\big)\\\nonumber
&+W\big(24\varpi\mathfrak{B_c}+Ze^{2r^2W}\big(-6\varpi\mathfrak{B_c}+32\pi\mathfrak{B_c}r^2-1\big)-2\big)-16
\varpi\mathfrak{B_c}r^4W^3Ze^{2r^2W}\big)\bigg],\\\nonumber
P_\bot&=\bigg[2\big\{4\pi\big(r^2WZe^{2r^2W}+1\big)^2+\varpi
W\big(-Ze^{2r^2W}+4r^2W+2r^4W^2Ze^{2r^2W}+3\big)\big\}\\\nonumber
&\times\big\{32\pi\big(r^2WZe^{2r^2W}+1\big)^3-\varpi
W\big(2r^2W\big(10Z^2e^{4r^2W}-23Ze^{2r^2W}-6\big)+17Ze^{2r^2W}\\\nonumber
&+32r^6W^3Z^2e^{4r^2W}+4r^4W^2Ze^{2r^2W}\big(Z^2e^{4r^2W}+6\big)-50\big)\big\}\bigg]^{-1}\bigg[W
\big\{\varpi  W \big(-24 \varpi  \mathfrak{B_c}\\\nonumber &-3 Z^2
(4 \varpi \mathfrak{B_c} -5) e^{4 r^2 W}+2 r^2 W \big(4 (8 \varpi
\mathfrak{B_c} +33)+2 Z^3 (11 \varpi \mathfrak{B_c} +7) e^{6 r^2
W}-Z^2\\\nonumber &\times(104 \varpi  \mathfrak{B_c} +89) e^{4 r^2
W}+4 Z (74 \varpi \mathfrak{B_c} -13) e^{2 r^2 W}\big)+2 Z (54
\varpi \mathfrak{B_c} -73) e^{2 r^2 W}\\\nonumber &+256 \varpi
\mathfrak{B_c} r^{10} W^5 Z^3 e^{6 r^2 W}+32 r^8 W^4 Z^2 e^{4 r^2 W}
\big(-4 (3 \varpi \mathfrak{B_c} +1)+\varpi  \mathfrak{B_c}  Z^2
e^{4 r^2 W}\\\nonumber &-2 Z (\varpi \mathfrak{B_c} -3) e^{2 r^2
W}\big)+8 r^6 W^3 Z e^{2 r^2 W} \big(4 (4 \varpi  \mathfrak{B_c}
-3)+5 \varpi \mathfrak{B_c} Z^3 e^{6 r^2 W}+2 Z^2 (4
\varpi\\\nonumber &\times\mathfrak{B_c} +5) e^{4 r^2 W}-4 Z (29
\varpi \mathfrak{B_c} -1) e^{2 r^2 W}\big)+4 r^4 W^2 \big(24 \varpi
\mathfrak{B_c} +Z^4 (4 \varpi  \mathfrak{B_c} +1) e^{8 r^2
W}\\\nonumber &+Z^3 (12 \varpi \mathfrak{B_c} +5) e^{6 r^2 W}-4 Z^2
(21 \varpi \mathfrak{B_c} +17) e^{4 r^2 W}+2 Z (66 \varpi
\mathfrak{B_c} +19) e^{2 r^2 W}+12\big)\\\nonumber &+152\big)-32 \pi
\big(r^2 W Z e^{2 r^2 W}+1\big)^2 \big(r^2 W \big(-4 (\varpi
\mathfrak{B_c} +1)+Z^2 (4 \varpi \mathfrak{B_c} +1) e^{4 r^2
W}-2\\\nonumber &\times Z (5 \varpi \mathfrak{B_c} +2) e^{2 r^2
W}\big)+e^{2 r^2 W} (Z-\varpi \mathfrak{B_c}  Z)+8
\varpi\mathfrak{B_c}  r^6 W^3 Z^2 e^{4 r^2 W}+2 r^4 W^2 Z e^{2 r^2
W}\\\nonumber &\times\big(\varpi  \mathfrak{B_c} \big(5 Z e^{2 r^2
W}-8\big)-2\big)-4\big)\big\}\bigg].
\end{align}
The adiabatic index corresponding to $\mathbb{L}_m=-\mu$ and
$\mathbb{L}_m=P_r$ has the form, respectively
\begin{align}\nonumber
\Gamma&=\bigg[3\big(16W^2XYr^2e^{2Wr^2}+1\big)\big\{128W^4\mathfrak{B_c}XYr^2e^{2Wr^2}\big(2XYe^{2Wr^2}\big(8\pi
r^2-\varpi\big)-\varpi r^2\big)\\\nonumber
&-32W^3XYr^2(2\varpi\mathfrak{B_c}+1)e^{2Wr^2}+8W^2XYe^{2Wr^2}\big(-10\varpi\mathfrak{B_c}+32\pi\mathfrak{B_c}r^2-1\big)+W\\\nonumber
&\times(8\varpi\mathfrak{B_c}
-1)+8\pi\mathfrak{B_c}\big\}\bigg]^{-1}\bigg[2W\big\{-\varpi\mathfrak{B_c}+1024W^4X^2Y^2r^4(2\varpi\mathfrak{B_c}-1)e^{4Wr^2}\\\nonumber
&+64W^3XYr^2e^{2Wr^2}\big(4XY(2\varpi\mathfrak{B_c}-1)e^{2Wr^2}-\varpi\mathfrak{B_c}r^2\big)+48W^2XYr^2(3\varpi\mathfrak{B_c}-2)\\\nonumber
&\times
e^{2Wr^2}+2W\big(4XY(5\varpi\mathfrak{B_c}-2)e^{2Wr^2}-3\varpi\mathfrak{B_c}r^2\big)-2\big\}\bigg],\\\nonumber
\Gamma&=\bigg[3\big(16W^2XYr^2e^{2Wr^2}+1\big)\big\{128W^4\mathfrak{B_c}XYr^2e^{2Wr^2}\big(2XYe^{2Wr^2}\big(8\pi
r^2-\varpi\big)-\varpi r^2\big)\\\nonumber
&+32W^3XYr^2(2\varpi\mathfrak{B_c}-1)e^{2Wr^2}+8W^2XYe^{2Wr^2}\big(-6\varpi\mathfrak{B_c}+32\pi\mathfrak{B_c}r^2-1\big)+W\\\nonumber
&\times(12\varpi\mathfrak{B_c}-1)+8\pi\mathfrak{B_c}\big\}\bigg]^{-1}
\bigg[2W\big\{-\varpi\mathfrak{B_c}+1024W^4X^2Y^2r^4(2\varpi\mathfrak{B_c}-1)e^{4Wr^2}\\\nonumber
&+64W^3XYr^2e^{2Wr^2}\big(4XY(2\varpi\mathfrak{B_c}-1)e^{2Wr^2}-\varpi\mathfrak{B_c}r^2\big)+48W^2XYr^2(3\varpi\mathfrak{B_c}-2)\\\nonumber
&\times
e^{2Wr^2}+2W\big(4XY(5\varpi\mathfrak{B_c}-2)e^{2Wr^2}-3\varpi\mathfrak{B_c}r^2\big)-2\big\}\bigg].
\end{align}

\end{document}